%% file: 0-main.tex
\documentclass[10pt,conference]{IEEEtran}
\IEEEoverridecommandlockouts
\usepackage{cite}
\usepackage{amsmath,amssymb,amsfonts}
\usepackage{algorithmic}
\usepackage{graphicx}
\usepackage{textcomp}
\usepackage{xcolor}
\usepackage{booktabs}
\usepackage{xspace}
\usepackage{amssymb,amsmath}
\usepackage{balance}
\usepackage{flushend}
\usepackage{subcaption}
\usepackage{tcolorbox}
\usepackage{multicol}
\usepackage{multirow}
\usepackage[export]{adjustbox}
\usepackage{hyperref}

\newboolean{showcomments}
\setboolean{showcomments}{false}
\ifthenelse{\boolean{showcomments}}
 { \newcommand{\mynote}[2]{
      \fbox{\bfseries\sffamily\scriptsize#1}
        {\small$\blacktriangleright$\textsf{\emph{#2}}$\blacktriangleleft$}}}
        { \newcommand{\mynote}[2]{}}

\definecolor{DarkOrange}{rgb}{0.8,0.3,0.0}
\definecolor{DarkCyan}{rgb}{0.0, 0.55, 0.55}

\def\BibTeX{{\rm B\kern-.05em{\sc i\kern-.025em b}\kern-.08em
    T\kern-.1667em\lower.7ex\hbox{E}\kern-.125emX}}
\begin{document}
% CB-NatStats 
% pred2nat
% CB2nat
% CodeBERT2nat
\title{{\sc CodeBERT-nt}: code naturalness via CodeBERT}

\newcommand{\toolname}{{\sc CodeBERT-nt}\xspace}
\newcommand{\cosin}{CBnt\_cos\xspace}
\newcommand{\acc}{CBnt\_acc\xspace}
\newcommand{\conf}{CBnt\_conf\xspace}
\newcommand{\VDA}{$\hat{\text{A}}_{12}$\xspace}
\newcommand{\VD}{Vargha and Delaney $\hat{\text{A}}_{12}$\xspace}

\newcommand{\changing}[1]{{#1}}
\author{Ahmed Khanfir, Matthieu Jimenez, Mike Papadakis and Yves Le Traon\\
	\normalsize University of Luxembourg, Luxembourg, Luxembourg\\
	\normalsize ahmed.khanfir@uni.lu, matthieu.jimenez@uni.lu, michail.papadakis@uni.lu, yves.letraon@uni.lu\\
	
}
%\newcommand{\changing}[1]{{\color{red} #1}} % remove the color here when done
%\and
%\and
%\and
%\and
%\and

\maketitle

\begin{abstract}
% The naturalness of a source-code is the 
% The naturalness of source-code is widely used by researchers in software engineering as an established static-analysis outcome-indicator of unusual or exotic programming practices.
% Much research on the software engineering field relies on the naturalness of source-code as an established static-analysis outcome-indicator of unusual or exotic programming practices.
% Recent static-analysis directions of the source-code has proven that unnatural source-code -> unusual or exotic programming practices he naturalness of source-code is widely used by researchers in software engineering as an established static-analysis outcome-indicator of unusual or exotic programming practices.
% Recent research has proven that the analysis of source-code naturalness can detect unusual or bad programming practices.  

Much of software-engineering research relies on the naturalness of code, the fact that code, in small code snippets, is repetitive and can be predicted using statistical language models like n-gram. % especially when it comes to detecting unusual and bad programming practices. %, which are often considered as symptoms of bug-proneness and bugginess. 
%Considering source-code as a recurrent sequences of words, the naturalness of a given sequence is typically estimated from its likelihood to occur. 
%Typically, language statistical models like n-gram ones are used to estimate the naturalness of a code-sequence based on its likelihood of occurrence. 
%Typically, language statistical models -- like n-gram ones -- are used to estimate the so-called naturalness of code. 
Although powerful, training such models on large code corpus is tedious, time consuming and sensitive to code patterns (and practices) encountered during training. %, making them hard to generalize to unseen code/projects.
Consequently, these models are often trained on a small corpora and estimate the language naturalness that is relative to a specific style of programming or type of project. 
%New language models like CodeBERT, however,  
%The n-gram language statistical models for instance have been proven useful/accurate in a wide variety of natural language processing NLP applications including the inferring of language naturalness indicators 
%Although, 
%This likelihood is however linked and impacted by the choice of the model training set, and thus reflects a naturalness judgement that is relative/bound to a specific space/scope/style of programming.
%This limitation motivates us to investigate a scope-independent naturalness of source-code, independantly  aim at capturing a scope-independent naturalness aspect of source-code  
To overcome these issues, we propose using pre-trained language models to infer code naturalness. Pre-trained models are often built on big data, are easy to use in  an out-of-the-box way and include powerful learning associations mechanisms. % dataset, in inferring code naturalness.
%investigate the capability of CodeBERT, a pre-trained language model on a large-scale projects dataset, in inferring code naturalness.
%To overcome the training set choice, we propose \toolname{} a novel code naturalness analysis approach/method that leverages the knowledge gained by CodeBERT -- a language model trained on a large scale dataset of projects -- to estimate the naturalness of source-code.
Our key idea is to quantify code naturalness through its predictability, by using state-of-the-art generative pre-trained language models. To this end, we infer naturalness by masking (omitting) code tokens, one at a time, of code-sequences, and  checking the models'  ability to predict them. To this end, we evaluate three different predictability metrics; a) measuring the number of exact matches of the predictions, b) computing the embedding similarity between the original and predicted code, i.e., similarity at the vector space, and c) computing the confidence of the model when doing the token completion task irrespective of the outcome. We implement this workflow, named \toolname{}, and evaluate its capability to prioritize buggy lines over non-buggy ones when ranking code based on its naturalness. Our results, on \changing{2,510} buggy versions of \changing{40} projects from the SmartShark dataset, show that \toolname{} outperforms both, random-uniform and complexity-based ranking techniques, and yields comparable results (slightly better) than the %JavaParser- and UTF8-tokenization based 
n-gram models.

%and challenges the model to predict their original values, then evaluates the 
%. Based on the  evaluating the prediction results variation through   

%scope-independent our attention to the use of a recently proposed pre-trained model CodeBERT. 
%Trained on over 6 million projects, the model comes with a better/larger/sharper understanding/knowledge of the programming language usual practices. 
%\toolname{}, a  language-naturalness extraction technique.  
%In this work we propose \toolname{} a novel CodeBERT based static-analysis technique to estimate the language naturalness of source-code, independently from its containing scope. % scope-independent 
%Trained on over 6 million projects, the model comes with a better/larger/sharper understanding/knowledge of the programming language usual practices. 

%Our work.
%Our results.
\end{abstract}

%\begin{IEEEkeywords}
%Code Naturalness, CodeBERT, Pre-trained models
%\end{IEEEkeywords}

\input{1-introduction}

\input{2-background}

\input{4-rqs}
\input{5-setup}

\input{6-results}

\input{7-discussion}

\input{9-conclusion}

\bibliographystyle{IEEEtran}
\bibliography{./bibliography/sample-base}

%\balance

\end{document}

%% file: 1-introduction.tex
\section{Introduction}
\label{sec:introduction}

There is a large body of research demonstrating that code, in small snippets, is repetitive and thus predictable~\cite{onTheNatSoft2012}. This repetitiveness of code is typically captured by statistical language models, like n-grams~\cite{allamanis2013mining}, that determine the appearance likelihood of a token given its $n$ preceding (or subsequent) ones. By computing the likelihood of tokens, in a sequence, and by using cross-entropy~\cite{predAndEntropyEnglish1950,mathematicalTheoryOfComm1948}, we can estimate the appearance likelihood of a given token sequence. 

In essence, n-gram-based cross-entropy quantifies the conformance of a given code sequence to the sequences appearing within the code corpus that was used during the n-gram training process. 
This naturalness metric of code can be quite powerful since it can identify unusual code sequences, that may reflect code that is smelly~\cite{SHARMA2021110936,impactRefactoringOnCNat2019}, of low readability~\cite{readability2021,readability}, or simply a rare specific implementation instance~\cite{bugram2016}.

Although powerful, training such models on large code corpus is tedious, time consuming and sensitive to code patterns (and specificities) of the used projects. % during training.
Additionally, the models are sensitive to a number of meta-parameters such as tokenizers, smoothing techniques, unknown threshold and $n$ values~\cite{jimenez2018impact}. Consequently, these models are often trained on a small corpus and estimate the language naturalness that is relative to a specific style of programming or type of project~\changing{\cite{onTheNatSoft2012}}. Moreover, their generalization ability, in a cross-project fashion, when trained on large corpus remains unclear since no such evidence appear in the related literature.  

To overcome these issues, we propose using generative pre-trained language models, such as CodeBERT, to infer code naturalness. Pre-trained models have been shown to provide strong results on several code-related tasks such as code generation and translation. They are often built on big data, are easy to use in an out-of-the-box way and include powerful learning associations mechanisms that allow them to generalize well to unseen code and projects. 

Therefore, our key objective is to form a powerful and flexible -- easy to use -- tool for computing code naturalness that is applicable in a cross-project fashion. %We thus, aim at using pre-trained language models, such as CodeBERT, that are powerful, readily available and trained over large corpus of data.
However, due to the generative nature of these models, they do not allow the computation of cross-entropy like measures, i.e., the probability of the appearance of given tokens of code. 

To deal with this issue we view code naturalness as a predictability metric. Thus, we measure how well code tokens can be generated by the models, and use it to infer naturalness. In particular, we mask (omit) code tokens, one at a time, of code sequences, and check the models’ ability to predict them. We then define and investigate the use of predictability metrics, that can be computed based on CodeBERT, to infer code naturalness. To this end, we evaluate three different predictability metrics:

\begin{itemize} 
    \item counting the number of exact matches of the predictions.
    \item computing the embedding similarity between the original and predicted code, i.e., similarity at the vector space.
    \item computing the confidence of the model when doing the token completion task, regardless to the outcome. 
\end{itemize}

We implement this workflow, which we call \toolname, and evaluate its suitability in inferring code naturalness. We also investigated the use of different aggregation methods on top of the above-defined token predictability metrics, % that we define above, 
such as minimum, maximum, mean, median and cross-entropy to infer the naturalness of code statements/lines as a whole thereby computing (n-token) naturalness and going beyond single tokens.

Since naturalness of code is a relative measure, i.e., it strongly depends on the models and training data used, and every metric involves its own specific building blocks, we cannot directly contrast the values of one metric over the others. To bypass this problem we compare the metrics
according to the relative performance differences on an end task, for which we have a solid ground truth. In particular, we evaluate the performance of our metrics  w.r.t. their ability to rank buggy lines of code.  

Previous work has shown that unusual code is often linked with bug-proneness and bugginess, thereby making  n-gram-based cross entropy a tool capable to identify likely buggy code \cite{bugram2016,onTheNaturalnessOfBC2016}. This means that buggy lines tend to be unnatural (entropy is high), or at least more unnatural compared to the clean ones that tend to be natural (entropy is low). Moreover, empirical studies \cite{onTheNaturalnessOfBC2016}, have shown that code naturalness has a comparable performance to other static bug finding approaches such as PMD~\cite{copeland2005pmd} and FindBugs~\cite{hovemeyer2007finding}.

We, therefore, investigate the ability of \toolname to distinguish natural (clean) from unnatural (buggy) code. To do so, we need a solid ground truth of buggy and clean code. Hence we select the SmartShark dataset \cite{Trautsch2021}, which contains manually untangled buggy and fixed code versions. This allows us to know the buggy code lines for which we can perform our experiment. We thus, used 2,510 buggy versions from 40 projects and investigated the performance of \toolname and contrast it with that of typical baselines such as uniform-random and complexity-based rankings~\cite{ChekamPBTS20}. 

To further strengthen our study, we also compared our results with that of n-gram models when trained on an intra-project fashion (as typically performed in literature~\cite{jimenez2018impact,onTheNaturalnessOfBC2016}).
Precisely, for every subject bug, we split its corresponding project-version code into two sets of lines: 1) the training set which counts all neutral file lines (files not impacted by the fix-commit) and 2) the evaluation lines set which counts all lines from the remaining files.
We used recommended settings by Jimenez et al.~\cite{jimenez2018impact} -- {\it unknown threshold}$=1$, $n=4$ with {\it KN} as smoother -- and two fundamentally different tokenizers; 1) {\it UTF8} which tokenize the code as a natural-language text including the comments and javadoc  and 2) {\it Java Parser}~\cite{JavaParser} which tokenize the code according to the java language grammar, excluding comments and documentation. 
%, settings, intra project bla bla. 
%We have created two n-gram models per subject bug, using the non-buggy file-lines as training-corpus and used them to measure the cross-entropy of the lines wihin the buggy files.
Finally, we compared \toolname metrics with the cross-entropy measured by these models in terms of ranking the buggy lines. 
Our results suggest that our (inter-project) \toolname yields comparable results (slightly better) than the (intra-project) naturalness predictions of the n-gram models.

Overall, our primary contributions are: 

\begin{itemize}
    \item We demonstrate that pre-trained generative models like CodeBERT capture the language naturalness notion. We can infer the naturalness aspect of a source-code from the CodeBERT prediction results. 
    
    \item We introduce a novel approach to detect source-code naturalness that works in cross-project context and does not rely on the aspects of naturalness that are tight to a specific project. For instance, It does not require any further training or specific knowledge of the target project, to rank its lines by bugginess likelihood, thus can be easily used in the future as a baseline.
    
     \item We provide empirical evidence demonstrating that the proposed approach is comparable to the current state-of-the-art.
\end{itemize}

%% file: 2-background.tex
\section{Background and related work}
\label{sec:background}

\subsection{Code-naturalness}
%Hindle et al.~\cite{onTheNatSoft2012} 
Estimating the source-code naturalness and predictability has been widely investigated.
Based on the fact that code is repetitive, the measure of naturalness was adopted on programming languages~\cite{onTheNatSoft2012}, using fundamentally the same techniques as for natural languages. %  similarly to 
Notably, Language models and particularly n-grams are the most popular medium of computing this measure.
%From the time Shanonn ~\cite{predAndEntropyEnglish1950}
N-gram models approximate the naturalness of a sequence of tokens based on its occurrence likely-hood, estimated relatively to the sequences observed in the training set.
This probability follows a Markov chain conditional probability series, where the probability $P(t)$ of a token $t$ to occur depends on the $n-1$ preceding tokens. 
To highlight irregular sequences (low probabilities of occurring), the naturalness of a sequence is expressed as a cross-entropy~\cite{predAndEntropyEnglish1950,mathematicalTheoryOfComm1948} which is computed by aggregating the logarithm of its token probabilities as follows:
\begin{equation}
\text{H}(S) = - \frac{\sum_{i=1}^{m} \log(P(t_i| t_{i-n+1}...t_{i-1}))}{n} ,
\end{equation}
where $n$ denotes the order of the n-gram model, $\{t_1,..., t_m\}$ the set of $m$ tokens forming the sequence $S$ and $P(t_i| t_{i-n+1}...t_{i-1})$ the probability $P(t_i)$ knowing $t_{i-n+1}...t_{i-1}$.
Consequently, n-gram models attribute high entropy values to unusual (not natural) code relatively to regular (natural) code.
To avoid assigning a zero probability to tokens that do not appear in the model vocabulary (which is often the case of variable names), n-gram models replace every token occurring less than $k$ times by a placeholder \texttt{<UNK>} and attributes a not zero probability to it, where $k$ and \texttt{<UNK>} are usually called the {\it unknown threshold} and {\it unknown word}~\cite{allamanis2013mining,CHEN1999359,jimenez2018impact}.
Similarly, to deal with unseen sequences of tokens in the training set, several smoothing techniques have been proposed and evaluated, among which Kneser Ney (KN)~\cite{KN1995} and Modified Kneser Ney (NKM)~\cite{CHEN1999359} are the best performing {\it smoothers}~\cite{CHEN1999359,onTheNatSoft2012}, with an eventual advantage for the latter one~\cite{jimenez2018impact}.
%\mj{we are going from  description of ngram to use of naturalness, perhaps this last part can be move up so that we have 1/ code naturalness, 2/ ngram, 3/ codebert.}
%\ahmed{hmmm it's use of n-gram too: 1) Baishakhi et al. in finding bugs 2) Jimenez et al. on sensitivity of n-grams. but feel free to refactor it if you prefer so.}
Among the potential applications of the naturalness of code property, 
Baishakhi et al.\cite{onTheNaturalnessOfBC2016} have shown empirical evidence that buggy lines are on average less natural than not-buggy ones and that n-gram entropy can be useful in guiding bug-finding tasks at both file- and line-level. %and that bug-fixing tend to increase the naturalness of code ~.
Jimenez et al.~\cite{jimenez2018impact} %carried-out a similar study with the aim to investigate the impact 
evaluated the sensitivity of n-gram w.r.t. its parameters and code tokenization techniques, via a file level naturalness-study. % and  parameters. %on the naturalness estimations accuracy. 
Their results confirmed Baishakhi et al. findings and provided recommendations on the best n-gram configurations for naturalness based applications, including the differentiation between buggy and fixed code.

% These finding were further confirmed by the study of

%Same as natural languages, programming languages are repetitive, thus their naturalness and predictability can be estimated by statistical-language-models 
%Based on the repetitiveness of source-code ,          

\subsection{CodeBERT}
Microsoft has recently introduced CodeBERT~\cite{DBLP:conf/emnlp/FengGTDFGS0LJZ20}, a bimodal pre-trained language model that supports multiple NL-PL (Natural Language and Programming Language) applications such as code search, code documentation generation, etc. 
The model was developed with a Multilayer Transformer~\cite{vaswani2017attention} architecture, which is adopted by the majority of large pre-trained models like BERT~\cite{devlin2018bert}. 
It has been trained on a large-scale dataset counting over 6 million projects from GitHub in 6 different programming languages, including Java. 
To ensure its bimodal functionalities, CodeBERT was trained towards a hybrid objective function (based on replaced token detection) in a cross-modal style, based on bimodal NL-PL data (precisely, source-code paired with its documentation) and unimodal data (including NL and PL sequences).
In this work, we incorporate its Masked Language Modeling (MLM) functionality~\cite{codebertweb} in our experiments pipeline, in order to study the possibility of inferring code naturalness from the CodeBERT prediction results.

%% file: 4-rqs.tex
\section{Research Questions}
\label{sec:rqs}

Our main objective is to form a flexible and easy-to-use tool for computing naturalness of code in a cross-project fashion. We thus, aim at using pre-trained generative language models, such as CodeBERT, that are powerful, readily available and trained over very large corpora. Since naturalness is a metric of how predictable a piece of code is and such pre-trained %generative
models are trained to predict code tokens, given their surrounding code, there should be a way to exploit them and compute code naturalness. However, since these models are generative, they do not compute the probability of the appearance of given tokens of code thereby obviating the naturalness calculation. To this end, we investigate the use of potential predictability metrics, that can be computed based on CodeBERT, and allow quantifying the predictability of code, which in fact represents code naturalness.  

Since naturalness of code is a relative measure, i.e., it strongly depends on the models and training data used, we cannot directly contrast the values of one metric over the others. We therefore, contrast them based on their performance on an end task, for which we have a solid ground truth. In particular, we evaluate the performance of our metrics w.r.t. the naturalness of bugs hypothesis that states: ''{\it unnatural code is more likely to be buggy than natural code}''~\cite{onTheNaturalnessOfBC2016}. This means that a good naturalness metric should be capable of distinguishing natural (clean) from unnatural (buggy) code.   
%Our main objective is to evaluate the relation between the naturalness of source code and CodeBERT predictions.
%is to assess if the naturalness of the code source has an impact on the CodBERT predictions and if we can infer the naturalness of a source code from the CodeBERT prediction results.
%Based on the source-code naturalness hypothesis that states that "unnatural code is more likely to be buggy", we aim at evaluating the effectiveness of CodeBERT on ranking source-codes by their suspiciousness to be buggy.  
%how well the variance of the prediction performance of CodeBERT is related tof the  that a source-code is more likely to be buggy or not from the CodeBERT predictions and based on the naturalness hypothesis; "not natual code is more likely to be buggy".
Hence, we ask:
%\mj{missing link with rq here, something like first we need to come up with way to evaluate natural based on codebert output}

\begin{description}
\item[RQ1] \emph{(Metrics and aggregation methods):} Which metrics and line-level aggregation methods lead to the best segregation between buggy and not buggy lines? 
\end{description}

To answer this question, we check if any pairs of metric-aggregation discriminate better the buggy from the clean lines. To do so, we rank and compare the results of both - ascendant and descendant - sorting orders on all possible pairs to determine the one that discriminates best. %This comparison depicts in which sorting direction the metric-aggregation pairs ranks the buggy lines better.
Using these matching sorting orders for every pair, we select the most effective aggregation method for every metric in terms of ranking buggy lines. 

However, getting the best ranking does not necessarily mean that our metrics perform well. We thus, turn our attention to the significance of these results and contrast them with some obvious baselines, such as the random order and source-code complexity order (ranking the most complex lines first). Random order offers a sanity check for coincidental results, and complexity offers an unsupervised baseline~\cite{surveyDefectPrediction2020}, i.e., shows that our metrics do not simply measure complexity instead of naturalness. Therefore, we ask:

%the contribution of the inferred information from CodeBERT on ranking the buggy lines. Hence we ask:

\begin{description}
\item[RQ2] \emph{(Comparison with random and complexity based rankings):} How does CodeBERT naturalness compares with random and source-code complexity, in terms of buggy lines ranking? 
\end{description}
To check whether the naturalness captured by CodeBERT is useful, we compare %its ranking performance with two baseline ranking techniques:
with 1) a uniform random selection based technique and 2) a code-complexity based one, where the complexity is measured in terms of number of tokens and the most complex lines are considered more likely to be buggy. To make a fair comparison, we compare the techniques on ranking the same subject lines and the same buggy versions, which have each at least one buggy line. 

%The results answering this question give evidence that \toolname{} is effective in capturing source-code naturalness. 
%Although, we note that it leverages a pretrained model on a large scale dataset of open-source projects to execute this task. 
%This fact may give it a significant advantage when compared to basic approaches such as the random or the complexity based one. 
%Hence, it is more interesting to compare the proposed approach results with more enhanced techniques. 
As the above-mentioned baselines are relatively weak, we also compare our metrics with n-gram models that were originally used to show the naturalness of buggy code hypothesis. This leads us to the following question:

\begin{description}
\item[RQ3] \emph{(Comparison with n-gram):} How does CodeBERT compare with n-gram language  models, in terms of lines ranking by bugginess?  
\end{description}

We answer this question by ranking the subject lines based on their cross-entropies measured by n-gram models trained on the source-code of the same version of the considered project. 
In fact, for every considered bug we create two n-gram models (using each of UTF8 and Java Parser tokenizers) trained on the source code of the files that has not been changed by the fix patch. Then, we use these models to calculate the cross-entropies of the subject lines.
Finally, we compare the ranking results by the n-gram models with our proposed approach, the same way as we did in RQ2. 

Obviously, the RQ3 comparison does not aim at answering the question of which model is best, but to show a relative performance of the models. This is because the training corpora of the models are significantly different with each one having significant associated advantages. In particular, CodeBERT has an advantage of being trained on a much larger corpus  than the n-grams. However, n-grams have the advantage of operating on an intra-project fashion, which is significantly more powerful\changing{~\cite{surveyBigCodeNat2017}} than the inter-project cases, as the one of CodeBERT. %\ahmed{I found one survey on that topic :D }

%Up to this point, the comparisons to answer the questions RQ2 and RQ3 rely on ranking the buggy lines among the business-logic source-code. Although, the answers provide evidence that we can infer source-code bugginess leveraging CodeBERT predictions which is important and demonstrates the potential of our approach, it does not necessarily mean that such claims hold when considering all source-code lines.
%Therefore, we seek to investigate:

%the answers to the posed questions provide evidence that we can infer source-code bugginess leveraging CodeBERT predictions. Although, this is important and demonstrates the potential of our approach, it does not necessarily mean that . Therefore, we seek to investigate:

%\begin{description}
%\item[RQ4] \emph{(All lines comparison):} How does CodeBERT compare with state-of-the-art approaches when considering all source-code lines, in terms of lines ranking by bugginess?
%\end{description}

%To answer this question we check whether the ranking results of CodeBERT remain relatively promising even when considering the buggy lines outside of business-logic source-code portion. 

%The difference of RQ4 from the other RQs is that in RQ4, we rank all the lines of the files changed by the fix commits and we normalise by the total number of these lines instead of the business-logic lines only.    

%% file: 5-setup.tex
\section{Experimental Setup}
\label{sec:setup}
%\ahmed{define shorter and consistent names for the metrics and aggregations. They should be used in all tables and figures}
\subsection{Dataset \& Benchmark}

%To answer to our research questions, we conduct an empirical study to assess the possibility of inferring the naturalness of source-code via CodeBERT prediction.
%On the absence of natural and unnatural code datasets, we base our comparison on the capability to extract bugginess aspects from non naturalness of source-code.
%Therefore, we used 
To perform our study, we use %the buggy versions from 
a recently crafted dataset of commits; SmartSHARK~\cite{Trautsch2021}. %that includes manual analysis of buggy code. 
%As there is no available source-code database of natural and not natural database, we target he
This dataset is distributed as a MongoDB database describing commit-details extracted from multiple open-source python and java projects. 
A specificity of Smartshark is that its data are manually refined, i.e., the commits are untangled and the fix validated, hence reducing the threat of noisy data.%This dataset is distinguished by its commit-untangling and fix-commits-validation. 
%In fact, the authors manually untangled the commits and categorised them into refactoring and issue-fixing ones.
% This feature reduces the potential threat of considering non-buggy lines as buggy.

In our study, we use the issue-fixing commits of the 40 available java projects the dataset includes.  
%We leverage this feature to ensure an additional check that our  
%The specific feature of this dataset is that the commit changes have been untangled and split into rafactoring and issue-fixing changes.
%Thsi fi.

\subsubsection{Buggy versions selection} 
For most of the issues, SmartSHARK provides one or more related commits, among which the fixing-commits are labeled as validated fix-commits.
Thus, to build our bugs dataset, we exclude all issues having no corresponding fix-commit, or having multiple fix-commits. %\mj{explaining the latter might be good, see comment}\ahmed{cool, i uncommented it.}
While the first case is straight forward, the latter is applied to further reduce noise in the data as pinpointing bugs origin from multiple fixing commits is harder and more error prone.
This way, we obtain one fix-commit per issue where the changed lines form the complete set of buggy lines.
Then, as we focus on Java source-code naturalness, we exclude the issues whose fixes are not involving java files, e.g., configuration files.
Finally, we map the resulting issues to their related buggy version, i.e., the project version preceding the fix commit. 

\subsubsection{Lines subject to study} 
\label{subseq:lines_selection}
For every considered issue, we retrieve the files changed by the patching commit, among which we mark as buggy any line changed by the fixed commit and define as neutral ones the remaining lines in these files.
%Furthermore, to reduce the computing cost of CodeBERT predictions we have run it only on the source-code portions that are responsible for the business-logic and skipped the lines that are unlikely to be the main cause of a bug, such as the import lines, the class definitions, the method signatures and the fields declarations.
%To better observe the difference of trends between
% To avoid any bias that can be caused by the 

%Then to reduce the impact of other ranking-bias such as  
Then to ensure that the observed ranking performances are related to code naturalness extraction and not to any non-uniform distribution of the buggy-lines over the code, we exclude all lines that are not part of the business-logic of the program like the imports, the fields declaration, the brackets, etc. We do so to focus on logic types of faults, which are unlikely to be caused by non-business-logic statements as also suggested by the work of Rahman et al.~\cite{natSoftwareRevisited2019}. Moreover, the naturalness of these lines is unlikely to be insightful. 

Additionally, this allows us to exclude any interference between comparing the ranking-performance by naturalness and other forms of rankings related to the buggy lines unequal distribution between business-logic and not business-logic related source-code portions. %\mj{not sure to get this last sentence}\ahmed{buggy lines are often in business-logic code: it's explained in the next sentences.}
In fact, close to half of the considered bugs see all of their buggy lines located within the business-logic code, while only \changing{10\%} of the bugs see all of their buggy-lines outside of it. %\mj{likely a negation missing here, overall it is hard to understand those numbers, We could just emphasize on the fact that only couple of issues are left behing due to lack of business logic modification, ie the last sentence.} \ahmed{something is broken is this sentence. i think we should leave these numbers. though we can explain them better:
%+ almost 1/2 dataset: all buggy lines are business logic. + 10\% dataset : all buggy lines are outside of business-logic + in 90\% of dataset bugs: median of almost 90\% buggy lines are in business-logic + business logic lines form 60\% of the lines.}
Unaccounting for those \changing{10\%}, we observe a median of \changing{90\%} of buggy lines located within the business-logic code per bug, while this type of line only amounts to \changing{60\%} of the lines overall. %\mj{fine with this ?}
%Whereas, in the remaining \changing{90\%} portion of the dataset, a median of almost \changing{90\%} of the buggy lines is part of the business-logic code, whose lines -- the business logic related ones -- form less than \changing{60\%} of all lines.    

To sum-up, by excluding these lines we exclude \changing{257} bugs out of \changing{2510} to end up with a final dataset counting \changing{2253} buggy versions, represented by a set of business-logic buggy and not buggy lines.

\subsection{Experimental Procedure}

\subsubsection{CodeBERT-based naturalness metrics}

\begin{figure*}[t]
\centering 
    %\vspace{-1.0em}
    \includegraphics[width=\textwidth]{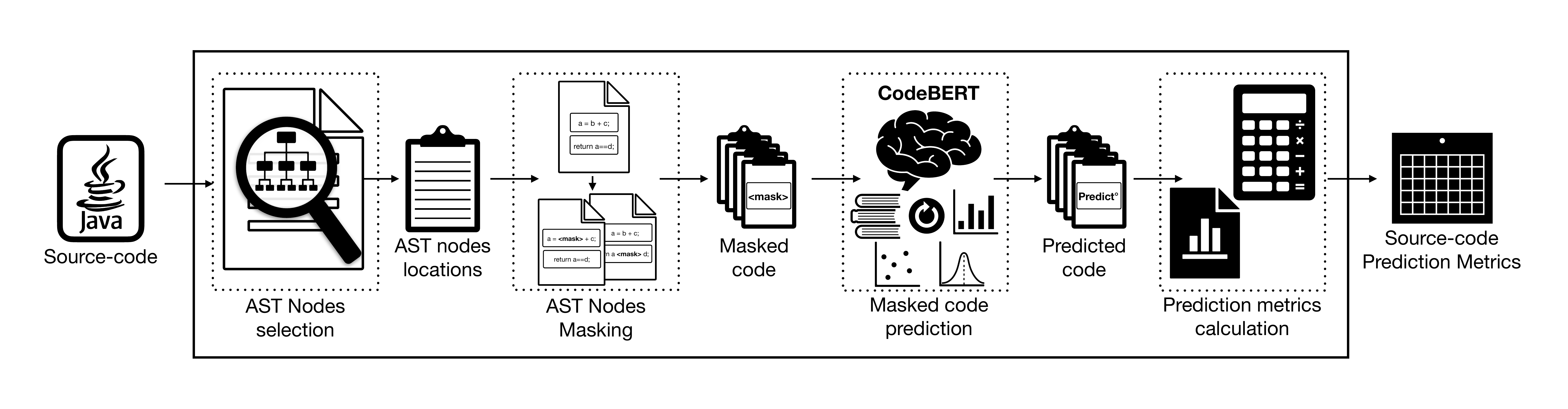}
    % \vspace{-3.5em}
    \caption{The \toolname source-code metrics calculation workflow.}
    \label{fig:pipeline}
    \vspace{-0.8em}
\end{figure*}

%We propose a novel static source-code analysis technique which leverages CodeBERT prediction variance to infer the variance of naturalness of a given input source-code. 
To infer the naturalness aspect of source-code from CodeBERT predictions we propose a 3-steps process that we name \toolname. 
%We illustrate an overview of its workflow in the Figure~\ref{fig:pipeline}.
\toolname operates in 3 steps as illustrated in the Figure~\ref{fig:pipeline}:

For every selected buggy file, the tool starts by parsing the Abstract Syntax Tree (AST) and extracting the interesting nodes to mask, excluding the language-specific tokens %\mj{control flow and import related token ?} 
such as control flow and import-related tokens, e.g., \texttt{if}, \texttt{else}, \texttt{for}, \texttt{while}, \texttt{import}, etc. 
Then, for every line, the tool iterates over the selected set of nodes and replaces each node's content by the placeholder \texttt{<mask>}, thus, generating one masked version of the input code per selected node.

Next, every masked version is tokenized into a vector of tokens using the CodeBERT tokenizer, and shrink to only encompass the masked token and its surroundings, as the model encoder can only take up to 512 tokens.
% to fit below the maximum number of tokens allowed by the model encoder (512 tokens). 
Once the shrinking is done, the sequences of masked codes are fed to CodeBERT %, surrounded by their neighbouring source-code in order 
to determine the best fitting substitute for the mask placeholder.
By default, the model provides 5 propositions ranked by likelihood (also called confidence) to match the masked node original value. In our setting, we only consider the first proposition as we believe it to be the most naturalness-revealing one (we discuss this choice further in Section~~\ref{subsec:pred-per-token}) and compute the following metrics from it: 
%For each proposition, we compute two additional metrics in addition to the prediction confidence, and save them as tuple of three values:
\begin{itemize}
    \item \emph{\conf{}:} the prediction confidence score of CodeBERT. This metric represents a probability, thus is a floating number varying between 0 and 1, where the closer to 1 its value is, the more CodeBERT is confident about the prediction. 
    We believe that this metric may mirror directly the naturalness of code as it reflects how predictable and usual is the code, relatively to the code knowledge learned by the model through its training phase on a large scale code dataset. 
    Thus, low confidence score may imply low naturalness.
    \item \emph{\cosin{}:} the cosine similarity between the CodeBERT embeddings of the predicted and the original source-code. This metric has a float value varying between 0 and 1, where 1 implies an exact similarity between the two embeddings and 0 the absence of similarity. This metric is often used in NLP and has shown some interesting results in filtering unnatural sentences \cite{9793859}.  
    CodeBERT embedding is the encoded representation of the code in the latin-space, where every token is represented by 1 vector. 
    To calculate the cosine similarity between the embeddings, we start by concatenating their token-vectors into two vectors, one for each embedding,
    then we compute the cosine as follows: 
\begin{equation}
     \text{Cosine}(\mathbf{V}_o,\mathbf{V}_p) = \frac{\mathbf{V}_o . \mathbf{V}_p}{\|\mathbf{V}_o\| . \|\mathbf{V}_p\|},
\end{equation}    
   where $\mathbf{V}_o$ and $\mathbf{V}_p$ are the concatenated embedding vectors of respectively the original and predicted code.
    
    Our intuition is that, the less natural the code is, the more CodeBERT will have issues to notice small changes in it, i.e., a difference in a single node, and thus, the higher the cosine will be.
    Consequently, the high similarity between both embeddings -- of the original and predicted code -- may be a symptom of unnaturalness of the code. 
    \item \emph{\acc{}:} the accuracy of prediction (whether the predicted code matches the original one or not). This is a boolean metric where 1 is attributed to a matching prediction and 0 otherwise. 
    Intuitively, we believe that the more the code is natural, the more CodeBERT predictions are accurate.
\end{itemize}
Finally, each line is mapped with its prediction-scores forming a matrix $\mathbf{M} \in \mathbb{R}^{3\times n}$ where $n$ is the number of collected propositions in that source-code line. 
%\ahmed{add equations.}
%\ahmed{precise the names of the metrics that will be used later in the legends of the figures}

The AST parsing and node-location extraction part has been implemented in Java and uses Spoon~\cite{spoon}, a Java code-source analysis library. 
Whereas, the rest of the process has been implemented in python and uses the CodeBERT Masked Language Modeling (MLM) task to predict the masked codes. 
%\ahmed{add intuition of the metrics: embeddings cosine -> natural code: small changes have impact Vs not natural: sees 512 tokens almost similarly. score -> more confident when the code is natural. match original -> more often in natural code. }

%Consequently, we attribute to every line a tuple three values corresponding to each of these metrics, which is calculated by aggregating the values of its nodes predictions tuples.
%\ahmed{talk about the SOA approaches: how we trained the n-gram models and how we calculate complexity and random.}
%\ahmed{speak about how each of these has been calculated in the next subsection - with the related RQ comparison}

\subsubsection{n-gram ranking}
\label{experimental_procedure_n-gram}
To compare CodeBERT with n-gram models in terms of code naturalness aspect capturing, we proceed as follows:

For every considered issue, we train two n-gram models specific to that version of the project using two distinct tokenization techniques\footnote{We use \texttt{UTFLineTokenizer} and \texttt{JavaLemmeLineTokenizer} available in Tuna~\cite{TUNA2018} GitHub repository under \texttt{tokenizer/line/} (branch=master,repo=\url{https://github.com/electricalwind/tuna}, rev-id=44188e1) }: 
\begin{itemize}
    \item Java Paraser tokenizer (noted JP)  which tokenizes the code according to the Java grammar, and thus, discarding empty lines as well as java-doc and code-comment lines. 
    \item UTF8 tokenizer (noted UTF8), which operates on the full raw representation of the source-code. 
\end{itemize}
We name the created models based on their underlying tokenization technique, JP and UTF8 n-gram.
%These tokenizers operate on different code representations, Java AST for   
In the training phase, we use all the lines from the files that have not been changed by the fix commit, %-- assumed to be not buggy --,
then we use each of these models to attribute a cross-entropy score to the subject lines of the buggy files, counting buggy and neutral lines as detailed in Sub-section~\ref{subseq:lines_selection}.
Finally, we rank the lines according to their cross-entropy score in a descendant order such-as high values of cross-entropy are associated with less code naturalness and higher likelihood of bugginess. 

To run this experiment, we use the current version of the n-gram utilities library Tuna~\cite{TUNA2018} with one of the recommended configurations by Jimenez et al.~\cite{jimenez2018impact} for distinguishing buggy and fixed lines: 4 as n-order, 1 as unknown threshold and Kneser Ney smoothing (KN). 
Note that we use KN instead of the Modified Kneser Ney smoothing (MKN) because it is not suited for short sequences.%, as is the case when operating on the line-level of granularity, in contrast with the file-level one~\cite{jimenez2018impact,TUNA2018}. 

%First, we split every buggy version source-code lines in our dataset into a testing set counting the same lines from the c the and testing set of lines; where the 
%Then, with the correspoding using two distinct tokenizers based on two different source-code representations we create two models for each bug separately. 

%The training test counts the  two sepecific n-gram models for every studied buggy version and  
%n-gram\cite{paper--n-gram} is a well established technique to build language statitstical models. 
%\subsubsection{Random ranking}

%\subsubsection{Compelexity ranking}

\subsection{Lines ranking}

To assess the relevance of the information inferred from the CodeBERT predictions, we rely on its ability to rank the buggy lines before the supposed neutral ones.
As we do not have any labelling of "natural" and "unnatural" source-code dataset and based on the naturalness hypothesis which assumes that "not natural lines are more likely to be buggy", we believe that the prediction variation of CodeBERT under naturalness variation of the input source-code can be observed through its ranking of buggy and not buggy lines.

For every considered approach, metric and aggregation method considered in our experiments, we rank all the lines by bug first, then normalise the ranks by the total number of studied lines for that bug.
We report two ranking results per bug per approach:
\begin{itemize}
\item the first hit: corresponds to the rank of the first-ranked buggy line.
\item mean: corresponds to the mean ranks of all buggy lines.
\end{itemize}
To cut ties when multiple lines share the same score, we attribute the estimated rank by a uniform random selection. For instance, if we have 100 lines sharing the same rank, among which 3 lines are buggy, the random first hit rank will be equal to 25, while the mean rank of the 3 buggy lines will be 50.

%To determine if we can infer any relationship between the CodeBERT prediction results and the bugginess of a line, we run a source-code prediction experiment on all subject lines of the considered bugs as detailed in \changing{section XX}.
%For every line, we mask one token, pass the masked code to CodeBERT-mlm to fill the mask. Then for every CodeBERT prediction (mask filling proposition), we extract the following metrics:
%1) the cosine similarity of the embeddings of the original code (not masked) and the code predicted by CodeBERT,
%2) the prediction score that depicts how confident CodeBERT is that the code predicted is likely to be correct,
%3) and if the predicted token matches the original masked one.

%RQ1)
To check whether any of the 3 aforementioned metrics are impacted by the naturalness variance of the source-code (answer to RQ1), we generate one CodeBERT prediction only by masked token.
Then we aggregate the scores of each line' predictions by applying one of the following aggregation metrics: minimum, maximum, mean, median and entropy.
%Finally to have one score per line, we aggregate the scores of its generated predictions 
Where the entropy is calculated as the following:
%\ahmed{entropy equation}.
% -(np.sum(np.log10(floats)) / float(len(floats)))
\begin{equation}
    \text{H}(l,m) = - \frac{\sum_{i=1}^{n} \log(s_i)}{n}
\end{equation}

where $\{s_1, s_2, ..., s_n\}$ denotes the set of $n$ scores attributed to the line $l$ for the metric $m$. 

We then rank the subject lines by each of the metrics using the different aggregation methods and following both sorting directions: ascending and descending.
As most of the results are close to each other, we calculate the paired \VD ratios, to conclude which combinations are the best in terms of buggy lines ranking.  

To check whether the extracted naturalness information from CodeBERT predictions (answer RQ2), we compare the ranking results of the 3 metrics using their best performing aggregation method and sorting order, with the rankings results of random and complexity-based rankings.
Where, the complexity of a line corresponds to its number of Java Parser tokens. 
%mpare the buggy lines ranking by CodeBERT metrics with a Random ranking and a Complexity based ranking, where the line complexity is equal to the number of its java tokens. 
This is inspired from the study of Leszak et al. \cite{LESZAK2002173}, where complex source code has been proven to be more likely to be buggy.
For the random ranking, instead of rerunning the ranking multiple times, we simply used a basic probability calculation of the rankings.

Finally, to compare the effectiveness of \toolname{} with similar techniques in capturing code naturalness we compare its buggy-line rankings with the n-gram cross-entropy, measured as described in Sub-section~\ref{experimental_procedure_n-gram} (answer to RQ3).
For this comparison, we consider the \toolname{} rankings effectuated by its best performing metric, selected from the previous research questions.

% compare its ranking results with the ones of n-gram models. from its best tuple metric-aggregation method   
%RQ3)
%We compare the buggy lines ranking by CodeBERT with the ranking by cross-entropy of n-gram models.
%Precisely, for every considered issue, we train two n-gram models  specific to that version of the project, using two different tokenization techniques; a Java Paraser (noted JP) tokenizer and a UTF8 tokenizer (noted UTF8). 
%In the training phase we have used all the lines from the files that have not been changed by the fix commit, then we used each of these models to attribute a cross-entropy score to the subject lines.
%Finally, we rank the lines according to their cross-entropy score in a descendant order such-as high values of cross-entropy are associated with less code naturalness and higher likelihood of bugginess. 
%The ranking results are then compared with the ones obtained leveraging CodeBERT predictions.

%RQ4 or discussion)
%We compare the ranking of buggy lines obtained by CodeBERT with the state of the art techniques when targeting all the lines and not the subset of the business-logic ones. 
%For CodeBERT, whenever a buggy line is not covered (does not have any score),  we attribute to it the worst score (the last rank).

To have a better understanding on the differences significance we run statistical tests of Wilkixon and \VD on all of our comparison results.

\subsection{Threats to Validity}
\label{sec:Threats}
\input{7-Threats}

%% file: 7-Threats.tex
%\ahmed{threats from different versions of java, threats related to the use of different dependencies (spoon, java parse, etc.), issues in the dataset... -> extra filtering of the dataset + considering the same bugs and lines where both tools covered at least one buggy line + considering all bugs and lines... }

The question of whether our findings generalise, forms a typical threat to validity of empirical studies. To reduce this threat, we used real-world projects, real faults and their associated commits, from an established and independently built benchmark. Still, we have to acknowledge that these may not be representative of projects from other programming languages, domains or industrial systems. 

Other threats may also arise from the assumption that all changed lines by the fix commits are buggy lines. While our heuristics are the standard in the literature, we believe that this selection process is sufficient given that we have used a dataset where the fix commits have been manually untangled. %To overcome this threat, w
Additionally, we focus our study on the sole business logic lines, thus reducing further the risk of considering as buggy, lines irrelevant to the bug at hand.
%have also studied the approaches performance on the business-logic lines only, thus focusing on a subset of lines that are most probably responsible of causing the issue. 

Finally, our evaluation metrics may induce some additional threats. Our comparison basis measurement, i.e., comparing the ranking of source-code lines that has been trained on the same source-code's project with approaches that are agnostic and has been trained on multiple projects. It is hard to compare the advantage gained by training CodeBERT on a large number of projects, including eventually the projects in our dataset, against the advantage gained by the n-gram models when trained on a specific project source-code and predicting the naturalness of lines of that same project.

%% file: 6-results.tex
\section{Results}
\label{sec:results}

\subsection{RQ1: Metrics and aggregation methods}
\label{subsec:rq1}

%\ahmed{include cosine and match\_orig in RQ1b and probably the rest of the RQs}
%\ahmed{discuss the trends of cosine and match\_orig also - if any}.
%\ahmed{regenerate the table 1 with improved style}
%\ahmed{fill the missing numbers of A12.}

%\begin{table*}
 %   \caption{Paired Vargha and Delaney $\hat{\text{A}}_{12}$ effect size values of the buggy lines ranking  by different pairs of metrics-aggregation methods in ascendant and descendant order. cosine refers to the Cosine similarity between the embeddings of the original code and the predicted one, score refers to the CodeBERT-confidence that the predicted code is probably correct and the match\_org metric refers to whether the predicted code matches the original one.}
  %  \centering
   % \subcaption*{1st buggy line ranking}
    %\scalebox{0.75}{
%    \input{figurestables/sorting_12_df__rq1a__p_1h_cbmupy}
 %   }
  %  \subcaption*{average buggy lines ranking}
   % \scalebox{0.75}{
    %\input{figurestables/sorting_12_df__rq1a__p_m_cbmupy}
%    }
 %   \label{tab:RQ1A12}
%\end{table*}

\begin{table*}
    \caption{Paired \VD effect size values of the buggy lines ranking  by different pairs of metrics-aggregation methods in ascendant and descendant order. cos refers to \cosin (the Cosine similarity between the embeddings of the original code and the predicted one), conf refers to \conf (the CodeBERT-confidence that the predicted code is probably correct) and the acc metric refers to \acc (whether the predicted code matches the original one).}
    \centering
    \scalebox{0.75}{
    \input{manual_figures/table_a12_rq1_sorting}
    }
    \label{tab:RQ1A12}
\end{table*}

\begin{figure}[t]
\centering 
    %\vspace{-1.0em}
    \includegraphics[width=0.5\textwidth]{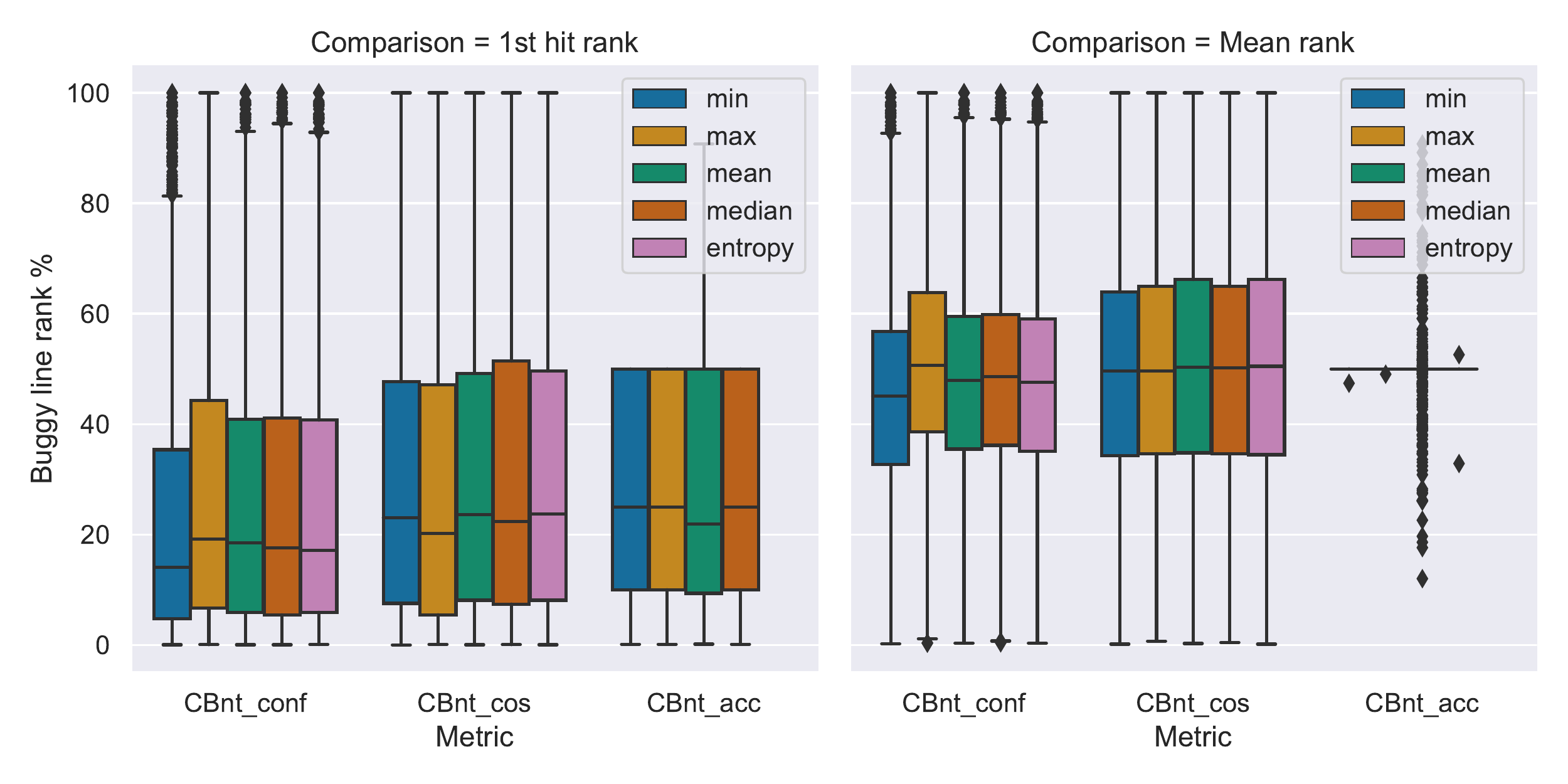}
    \caption{Buggy lines ranking using 1 prediction per token and the three available metrics with different aggregation metrics.
     The less CodeBERT is confident when predicting, the more likely the line is to be buggy.}
    \label{fig:RQ1a}
\end{figure}

% \begin{figure}[t]
% \centering 
%      \begin{subfigure}{0.5\textwidth}
%          \centering
%          \includegraphics[width=\textwidth]{figurestables/boxplot_df__rq1a__p_1h_cbmupy.pdf}
%         \vspace{-0.5em}
%          \caption{1st hit rank: 1st ranked buggy line.}
%          \label{fig:RQ1a_1h}
%      \end{subfigure}
%      \hfill
%      \begin{subfigure}[b]{0.5\textwidth}
%          \centering
%          \includegraphics[width=\textwidth]{figurestables/boxplot_df__rq1a__p_m_cbmupy.pdf}
%          \caption{Mean rank: mean rank of buggy lines.}
%          \label{fig:RQ1a_m}
%      \end{subfigure}
%      \hfill
%      \caption{Buggy lines ranking using 1 prediction per token and the three available metrics with different aggregation metrics.
%      The less CodeBERT is confident when predicting, the more likely the line is to be buggy.}
%     \label{fig:RQ1a}
% \end{figure}

To evaluate the \toolname metrics, we ranked the subject lines according to the aforementioned metrics. We start by calculating one score per line for every metric, by aggregating the scores of the predictions in that line. Then, we sort the lines according to their score in both orders - ascendant and descendant - and calculate the paired (by bug) \VD difference between both orders effectiveness in terms of attributing the lowest ranks to the buggy lines; more precisely the average ranking of the buggy lines and the smallest rank attributed to a buggy line, per bug. (Please refer to section~\changing{\ref{sec:setup}} for details) 

These results are depicted in Table~\ref{tab:RQ1A12} where values around 0,5 depict that the considered pair of metric-aggregation method does not bring any advantage in ranking buggy lines first, while values above 0,5 confirm an advantage for the ascendant sorting order and below 0,5 an advantage for the descendant one.
%From these results, we use the most suited sorting order for each pair of metric-aggregation and illustrate in 

Figure~\ref{fig:RQ1a} depicts the box-plots of the normalised rankings by number of lines of each bug. Interestingly, the prediction confidence score seems to be a good indicator of naturalness and thus of bugginess likelihood. Noticeably, low confidence scores are more often attributed to buggy lines than neutral ones. This explains why aggregating this metric by using the maximum value by line gives the worst guidance to our target, while sorting by the minimum score gives the best ranking over all considered pairs metric-aggregation.  

At the same time, we do not observe any relevant differences when ranking the lines by either the cosine of the embedding or the correctness of predictions. This is also confirmed by the Table~\ref{tab:RQ1A12} where most of the \VDA values are around 0,5 when using each of the aggregation methods on these two metrics. Nevertheless, we notice that the best method to rank the buggy lines by cosine similarity is via ranking the lines by their highest scores of similarity, where the high (max) values are considered as symptoms of unnaturalness. 
We also notice a trend that confirms the correlation between naturalness and code predictability, such as the lower the mean of correct predictions in a code is, the less natural is the code.

% be seen from the boxplots, the prediction confidence score can be a good indicator of naturalness and thus of buginess likelihood. 
%Noticeably, low confidence scores are more often attributed to buggy lines than not buggy ones. 
%This explains why aggregating this metric by using the maximum value by line gives the worst guidance to our target, while sorting by the minimum score gives the best ranking over all considered pairs metric-aggregation.  

By checking Table~\ref{tab:RQ1A12} we also see that the difference of ranking the lines by low confidence, from the other metric-aggregation methods is significant, \VD effect size values are more than \changing{0.6}, which are significantly higher to the rest of \changing{confidence aggregation rankings}. 
Less noticeable, the ranking by increase of the line's maximum embeddings cosine-simalirity and mean prediction accuracy yield respectively \VDA values of around \changing{0,52} and \changing{0,51} when compared to the other studied aggregation methods when applied on the same metrics.

\begin{tcolorbox}
\toolname{} can infer the naturalness of code through masked token predictions.
The unnatural information can be inferred the best from the decrease of prediction confidence (considering the minimum value per line), the increase of cosine similarity between the embedded original and predicted code (considering the maximum value per line) and the decrease of average prediction accuracy per line.
%Non-naturalness information can be inferred the best from the decrease of confidence of CodeBERT when filling masked code. 
\end{tcolorbox}

\subsection{RQ2: Comparison with random and complexity based rankings}

%\ahmed{fix figure legend}

\begin{figure}[t]
\centering 
    %\vspace{-1.0em}
    \includegraphics[width=0.5\textwidth]{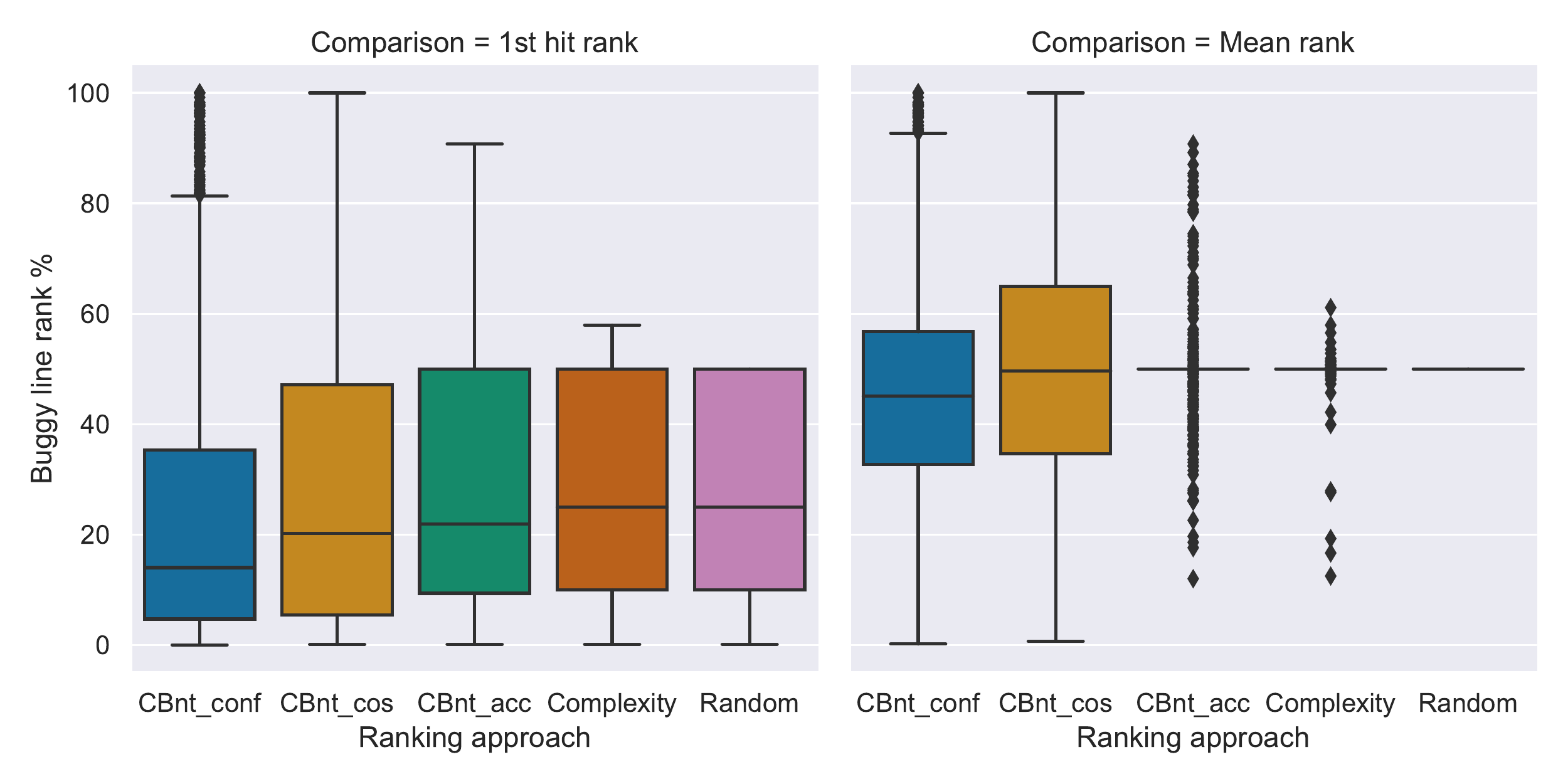}
    \caption{Comparison of the buggy lines rankings by CodeBERT, Random and Complexity (number of tokens by line).
     CodeBERT outperforms Random and Complexity in ranking buggy lines.}
    \label{fig:RQ2}
\end{figure}

% \begin{figure}[t]
% \centering 
%      \begin{subfigure}{0.5\textwidth}
%          \centering
%          \includegraphics[width=\textwidth]{figurestables/boxplot_df__rq2__p_1h_cbmupy.pdf}
%         \vspace{-0.5em}
%          \caption{1st hit rank: 1st ranked buggy line.}
%          \label{fig:RQ2_1h}
%      \end{subfigure}
%      \hfill
%      \begin{subfigure}[b]{0.5\textwidth}
%          \centering
%          \includegraphics[width=\textwidth]{figurestables/boxplot_df__rq2__p_m_cbmupy.pdf}
%          \caption{Mean rank: average rank of buggy lines.}
%          \label{fig:RQ2_m}
%      \end{subfigure}
%      \hfill
%      \caption{Comparison of the buggy lines rankings by CodeBERT, Random and Complexity (number of tokens by line).
%      CodeBERT outperforms Random and Complexity in ranking buggy lines.}
%     \label{fig:RQ2}
% \end{figure}

Figure \changing{\ref{fig:RQ2}} shows the distribution of the normalised rankings of the first ranked buggy line and the average rank of buggy lines by bug when using \toolname{} metric-aggregation pairs selected from the results of RQ1 (ascendant minimum \conf, descendent maximum \cosin and ascendant mean \acc{}), uniform random ranking and token-number-complexity descendant ranking.

Surprisingly, random and complexity lead to similar rankings with a small advantage for random. This observation implies that tokens-number-complexity does not capture well the code naturalness in the studied setup: naturalness on the line-level-granularity of the business-logic code.  

As can be seen from the boxplots, \toolname{} outperforms the baseline techniques in ranking the buggy lines first, using any of its three metrics.
In fact, except for a small portion of our dataset bugs, \toolname{} low-confidence performs the best in estimating the bugginess of the target lines, with respectively a 1st hit and mean buggy-line ranks lower in average by around \changing{6\%} and \changing{4.7\%} than the following ranks -- attributed by the cosine similarity -- and around \changing{11\%} and \changing{5\%} lower than the least performing ranking effectuated by \changing{uniform random}. 
This noticeable difference between the \toolname{} results with the two considered techniques, more specifically the low confidence \conf ranking, endorses the fact that \toolname{} can be considered as a comparison baseline and a new method for naturalness based tasks.%For a small portion of our dataset bugs, CodeBERT attributes high ranks to the buggy lines and performs worst than Random and Complexity. 

Although, the two remaining \toolname{} metrics, average accuracy \acc and high embeddings similarity \cosin, outperform both baselines and yield lower rankings than uniform random by respectively \changing{3\%} and \changing{5\%} in ranking the first hit buggy line, they perform however similarly to the baselines when comparing their mean ranking of all the buggy lines, with a small advantage of only \changing{0,4\%} for the ranking by embeddings similarity.
%without any improvement when ranking by accuracy and only \changing{0,4\%}. 

%Additionally, in Figure~\ref{fig:RQ2_m}, we see that ranking the lines by \toolname{} prediction accuracy lead to a low average ranking of the buggy lines and performs comparably to both baselines. 

\begin{table}[]
    \caption{Vargha and Delaney $\hat{\text{A}}_{12}$ of (\toolname low-confidence-metric rankings  compared to the other ones.) %\mj{is it really percent ?}
    }
    \centering
    \input{manual_figures/a12_table_rq2_}
    \label{tab:a12_rq2}
\end{table}

To validate this finding, we perform a statistical test (Wilcoxon paired test) on the data of Figure \changing{\ref{fig:RQ2}} to check for significant differences. Our results showed that the differences are significant, indicating the low probability of this effect to be happening by chance. As illustrated in Table~\ref{tab:a12_rq2}, the size of the difference is also big, with \toolname{} low-confidence yielding \VD values between \changing{0.58} and \changing{0.6} indicating that \conf ranks the buggy lines the best in the great majority of the cases. 
However, the \toolname{} accuracy and embeddings-similarity metrics outperform random respectively, in only \changing{51\%} and \changing{48\%} of the cases.

%\ahmed{add A12 wilkixcon table for RQ2}

\begin{tcolorbox}
\toolname{} metrics describe source-code naturalness more accurately than the baselines uniform-random-selection and tokens-count-complexity based rankings. 
\toolname{} low confidence -- \conf{} -- is the most effective metric and outperforms the uniform-random-ranking by \changing{11\%} in ranking the first hit buggy line and \changing{5\%} in ranking all the buggy lines, in average.
%, high embeddings similarity and average accuracy, outperforms the uniform-random-ranking by respectively \changing{11\%} , \changing{5\%} and \changing{3\%} in ranking the first hit buggy line and \changing{5\%} , \changing{0,4\%} and \changing{0\%} in ranking all the buggy lines, in average.   

\end{tcolorbox}

\subsection{RQ3: Comparison with n-gram}

\begin{figure}[t]
\centering 
    %\vspace{-1.0em}
    \includegraphics[width=0.5\textwidth]{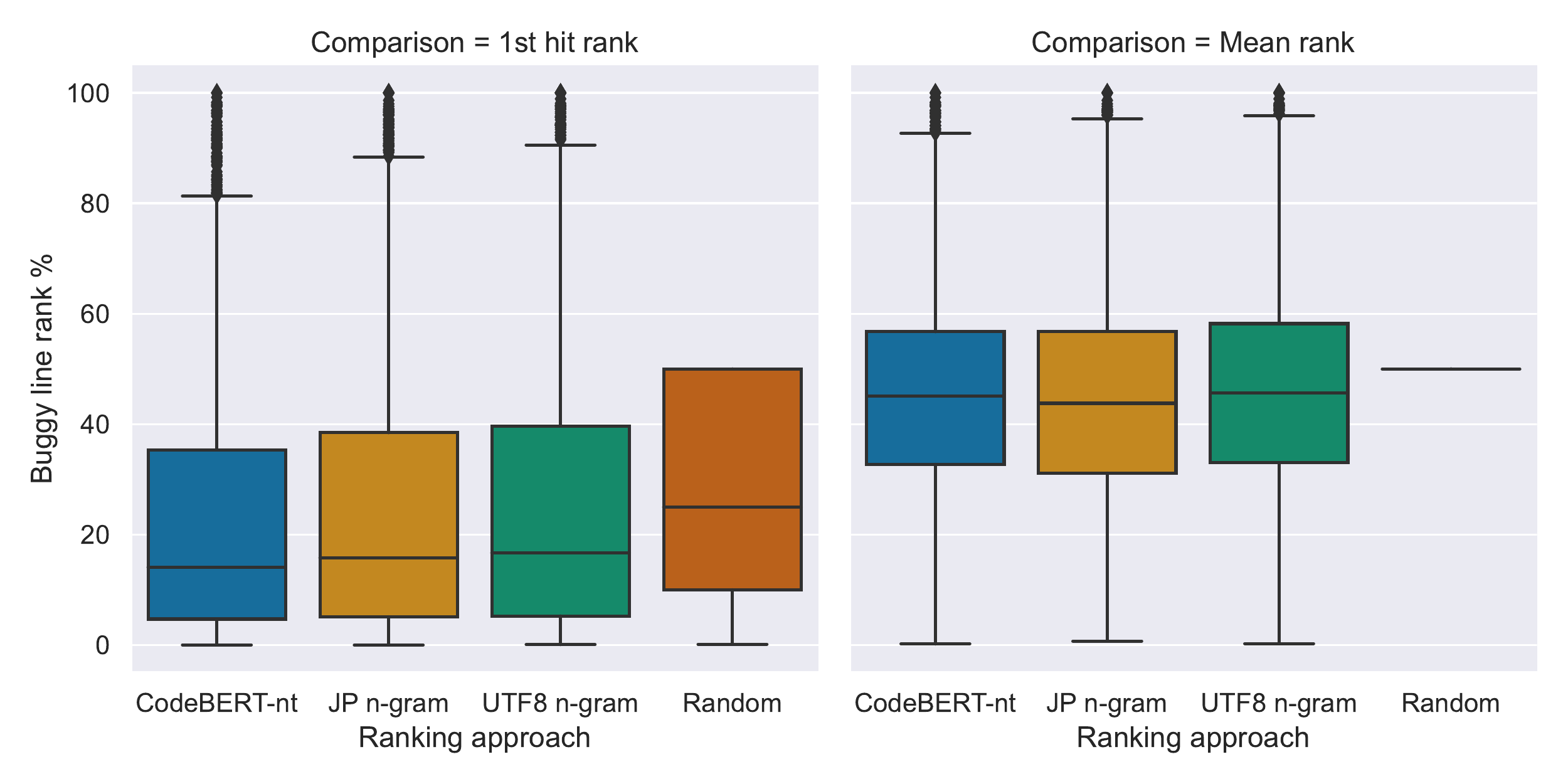}
    \caption{Comparison of the buggy lines rankings by CodeBERT, UTF8 n-gram and JP n-gram models (created respectively using UTF8 and Java Parser tokenizers).
     CodeBERT ranking is comparable to the n-gram models one.}
    \label{fig:RQ3}
\end{figure}

% \begin{figure}[t]
% \centering 
%      \begin{subfigure}{0.5\textwidth}
%          \centering
%          \includegraphics[width=\textwidth]{figurestables/boxplot_df__rq3__p_1h_cbmupy.pdf}
%         \vspace{-0.5em}
%          \caption{1st hit rank: 1st ranked buggy line.}
%          \label{fig:RQ3_1h}
%      \end{subfigure}
%      \hfill
%      \begin{subfigure}[b]{0.5\textwidth}
%          \centering
%          \includegraphics[width=\textwidth]{figurestables/boxplot_df__rq3__p_m_cbmupy.pdf}
%          \caption{Mean rank: average rank of buggy lines.}
%          \label{fig:RQ3_m}
%      \end{subfigure}
%      \hfill
%      \caption{Comparison of the buggy lines rankings by CodeBERT, UTF8 n-gram and JP n-gram models (created respectively using UTF8 and Java Parser tokenizers).
%      CodeBERT ranking is comparable to the n-gram models one.}
%     \label{fig:RQ3}
% \end{figure}

To answer this question, we train two n-gram models per buggy version of our dataset that we then  use to compute the cross-entropies of the subject lines corresponding to the corresponding bug, then we rank these lines according to the resulting values and reproduce the same analysis as in RQ2.   
We illustrate in Figure \changing{\ref{fig:RQ3}} the distribution of the normalised rankings of the first ranked buggy line and the average rank of buggy lines by bug when using \toolname{} low confidences ranking -- \conf{} --, the descendant cross-entropy ranking from a UTF8-tokenizer-based n-gram model and a JavaParser-tokenizer-based one. 
Additionally, we illustrate the random ranking in the boxplots as the simplest baseline for this task.

As expected, the three approaches outperform the uniform-random-ranking in most of the cases and yield very comparable results with a small advantage to \toolname{} on ranking the first buggy line over the n-gram techniques and a small advantage to JP n-gram in regards of the average rank of buggy lines. In both comparisons, UTF8 n-gram falls slightly behind these two latter techniques. 

\input{manual_figures/A12_RQ3}

To validate this finding, we performed a similar statistical test as in RQ2 on the data of Figure \changing{\ref{fig:RQ3}} and found that the differences with random are significant, while the differences between n-gram cross-entropy and \toolname{} low-confidence rankings are negligible. As illustrated in Table~\ref{tab:RQ3_a12}, the size of the \VDA differences are equal to \changing{0.518 and 0.468} between \toolname{} and JP n-gram models and \changing{0.536 and 0.502} for both reported rankings, meaning that both approaches yield comparable results, which confirms that \toolname{} can carry out code naturalness related applications with similar effectiveness as the statistical language n-gram models.
%\ahmed{add A12 values}

\begin{tcolorbox}
\toolname{} masked token prediction confidence indicates naturalness of bugs as accurately (slightly better) as program-specific n-gram models.
\end{tcolorbox}

%% file: manual_figures/table_a12_rq1_sorting.tex
\begin{tabular}{lrrrrrrrrrrrrrr}
\toprule
     \textbf{Metric} &  \textbf{conf min} &  \textbf{conf max} &  \textbf{conf mean} &  \textbf{conf median} &  \textbf{conf entropy} &  \textbf{cos min} &  \textbf{cos max} &  \textbf{cos mean} &  \textbf{cos median} &  \textbf{cos entropy} &  \textbf{acc min} &  \textbf{acc max} &  \textbf{acc mean} &  \textbf{acc median} \\
\midrule
\textbf{1st hit} &    0.6196 &    0.4993 &     0.5528 &       0.5542 &        0.4387 &   0.5084 &   0.4763 &    0.5016 &      0.4989 &       0.4980 &   0.5004 &   0.4996 &    0.5118 &      0.4996 \\
\textbf{Mean} &    0.6325 &    0.5009 &     0.5613 &       0.5530 &        0.4350 &   0.5138 &   0.4818 &    0.5027 &      0.5058 &       0.4980 &   0.5004 &   0.4996 &    0.5122 &      0.4996 \\
\bottomrule

\end{tabular}

%% file: manual_figures/a12_table_rq2_.tex
\begin{tabular}{lrrrr}
\toprule
\textbf{CBnt\_conf Vs} &  \textbf{CBnt\_cos} &  \textbf{CBnt\_acc} &  \textbf{Complexity} &  \textbf{Random} \\
\midrule
         \textbf{1st hit} &    0.578 &    0.609 &      0.605 &  0.607 \\
            \textbf{mean} &    0.565 &    0.619 &      0.620 &  0.622 \\
\bottomrule
\end{tabular}
\vspace{-2em}

%% file: manual_figures/A12_RQ3.tex
% \usepackage{multirow}
\begin{table}[]
\caption{Vargha and Delaney $\hat{\text{A}}_{12}$ of \toolname low-confidence-metric rankings compared to n-gram and Random ones.}

\begin{tabular}{|l|ll|llll}
\toprule
\multirow{2}{*}{\textbf{Approaches $\hat{\text{A}}_{12}$}} & \multicolumn{2}{c|}{\textbf{Random}} & \multicolumn{2}{c|}{\textbf{UTF8}} & \multicolumn{2}{c|}{\textbf{JP}} \\  
                         & \multicolumn{1}{l}{1st hit} & mean  & \multicolumn{1}{l}{1st hit} & \multicolumn{1}{l|}{mean}  & \multicolumn{1}{l}{1st hit} & \multicolumn{1}{l|}{mean}  \\ \midrule
\textbf{CBnt\_conf}               & \multicolumn{1}{l}{0.606}   & 0.621 & \multicolumn{1}{l}{0.536}   & \multicolumn{1}{l|}{0.502} & \multicolumn{1}{l}{0.518}   & \multicolumn{1}{l|}{0.468} \\ 
\textbf{JP}                       & \multicolumn{1}{l}{0.575}   & 0.630 & \multicolumn{1}{l}{0.559}   & \multicolumn{1}{l|}{0.588} &                              &           \multicolumn{1}{l|}{}                 \\ 
\textbf{UTF8}                     & \multicolumn{1}{l}{0.560}   & 0.602 &                              &                          \multicolumn{1}{l|}{}  &                              &       \multicolumn{1}{l|}{}                     \\ 
\bottomrule
\end{tabular}
\label{tab:RQ3_a12}
\end{table}

%% file: 7-discussion.tex
\section{Discussion}
\label{sec:discussion}

%\ahmed{number of bugs where each tool is better than the other}
%\ahmed{can we differentiate between these bugs?}

\subsection{Impact of interesting lines selection}

%This is important because it forms a fair base of comparison and excludes the impact of targeting business-logic from the contributions of the prediction metrics collection.

\begin{figure}[t]
\centering 
    %\vspace{-1.0em}
    \includegraphics[width=0.5\textwidth]{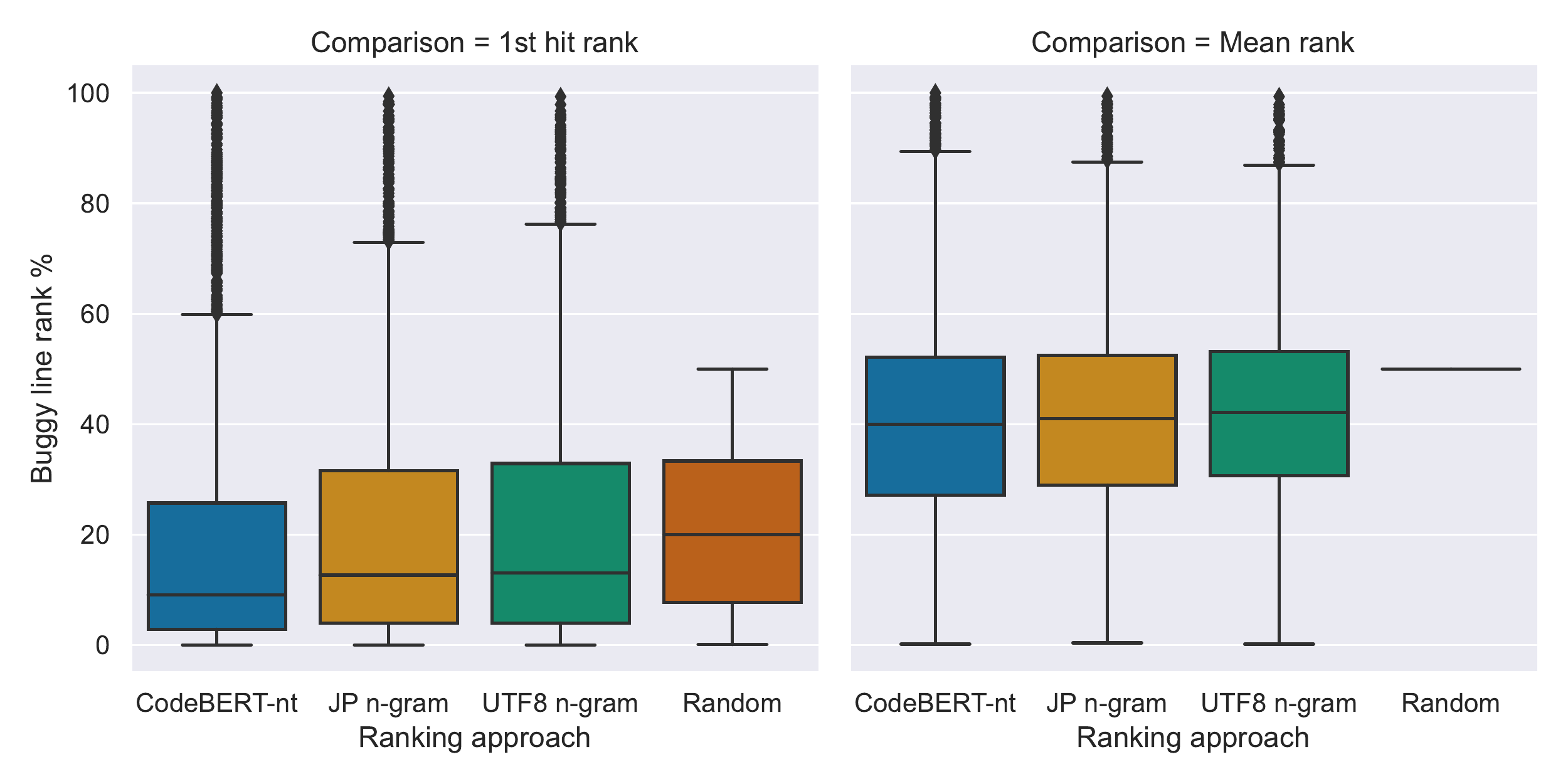}
    \caption{Comparison of the buggy lines rankings by CodeBERT, UTF8 n-gram and JP n-gram models (created respectively using UTF8 and Java Parser tokenizers) when ranking all lines.
     CodeBERT ranking is comparable to the n-gram models one and ranking the business-logic source-code lines first give it an advantage over the n-gram ranking.}
    \label{fig:RQ4}
\end{figure}

% \begin{figure}[t]
% \centering 
%      \begin{subfigure}{0.5\textwidth}
%          \centering
%          \includegraphics[width=\textwidth]{figurestables/boxplot_df__rq4__p_1h_tuna.pdf}
%         \vspace{-0.5em}
%          \caption{1st hit rank: 1st ranked buggy line.}
%          \label{fig:RQ4_1h}
%      \end{subfigure}
%      \hfill
%      \begin{subfigure}[b]{0.5\textwidth}
%          \centering
%          \includegraphics[width=\textwidth]{figurestables/boxplot_df__rq4__p_m_tuna.pdf}
%          \caption{Mean rank: average rank of buggy lines.}
%          \label{fig:RQ4_m}
%      \end{subfigure}
%      \hfill
%      \caption{Comparison of the buggy lines rankings by CodeBERT, UTF8 n-gram and JP n-gram models (created respectively using UTF8 and Java Parser tokenizers) when ranking all lines.
%      CodeBERT ranking is comparable to the n-gram models one and ranking the business-logic source-code lines first give it an advantage over the n-gram ranking.}
%     \label{fig:RQ4}
% \end{figure}

Our empirical results show evidence that \toolname{} can infer source-code naturalness yielding the same results as n-gram models and outperforming the uniform-random and code-complexity based techniques in attributing higher ranks to buggy lines when sorting source-code by naturalness.
These experiments have been driven on the same buggy versions source-code whose lines count at least one buggy line.
%However, it remains unclear whether these conclusions hold when considering all the bugs and lines from our dastaset. 
%Up to this point, to answer the previous questions we have compared performances of the proposed approach with the SOA ones on ranking the business-logic lines only. 
Precisely, we have excluded all the lines outside the business-logic source-code and kept only the bugs that counted at least one buggy line within their remaining lines.
%In this research question, we turn our attention to the impact of this lines selection step and reintroduce all the lines and bugs in our dataset.

To better understand the impact of this line selection step, we reintroduce all the bugs with their full source-code in our dataset and reproduce the same study as in \changing{RQ3}.
We have then attributed the worst rank to all unranked lines by \toolname{}, i.e., outside of the business-logic.
%Then we reproduce the same study as in the previous research question RQ3, with attributing the worst rank to all not ranked lines for CodeBERT.
This implies that for \toolname{}, the business-logic lines are ranked first by their min-confidence and the remaining lines are ranked after with a random uniform selection logic.
The n-gram approaches ranking is applied as described previously, the same way on all lines -- business- and non-business-logic related ones -- having each two cross-entropy values from every corresponding n-gram model.  
The ranking distributions are illustrated in Figure~~\changing{\ref{fig:RQ4}}. %and \ref{fig:RQ4_m}}. 

Although the results of the three approaches remain comparable, the trend is that for a noticeable portion of studied bugs, \toolname{} remains able to rank buggy lines better than the n-gram models. 
The difference is wider and more visible between the ranks of the first buggy line which can be seen in the left boxplot of Figure \changing{\ref{fig:RQ4}}.
Interestingly, we observe that ranking the business-logic lines with \toolname{} and the remaining lines with a uniform random ranking outperforms ranking all the lines (business- and not business-logic related ones) by their n-gram calculated cross-entropies. 
These results lead to the conclusion that the naturalness analysis of the non-business-logic lines do not contribute with useful information to the considered ranking tasks, but instead alters its results when attributing a higher rank to the targeted lines.

\begin{figure}[t]
\vspace{1.5em}
\centering 
    %\vspace{-1.0em}
    \includegraphics[width=0.5\textwidth]{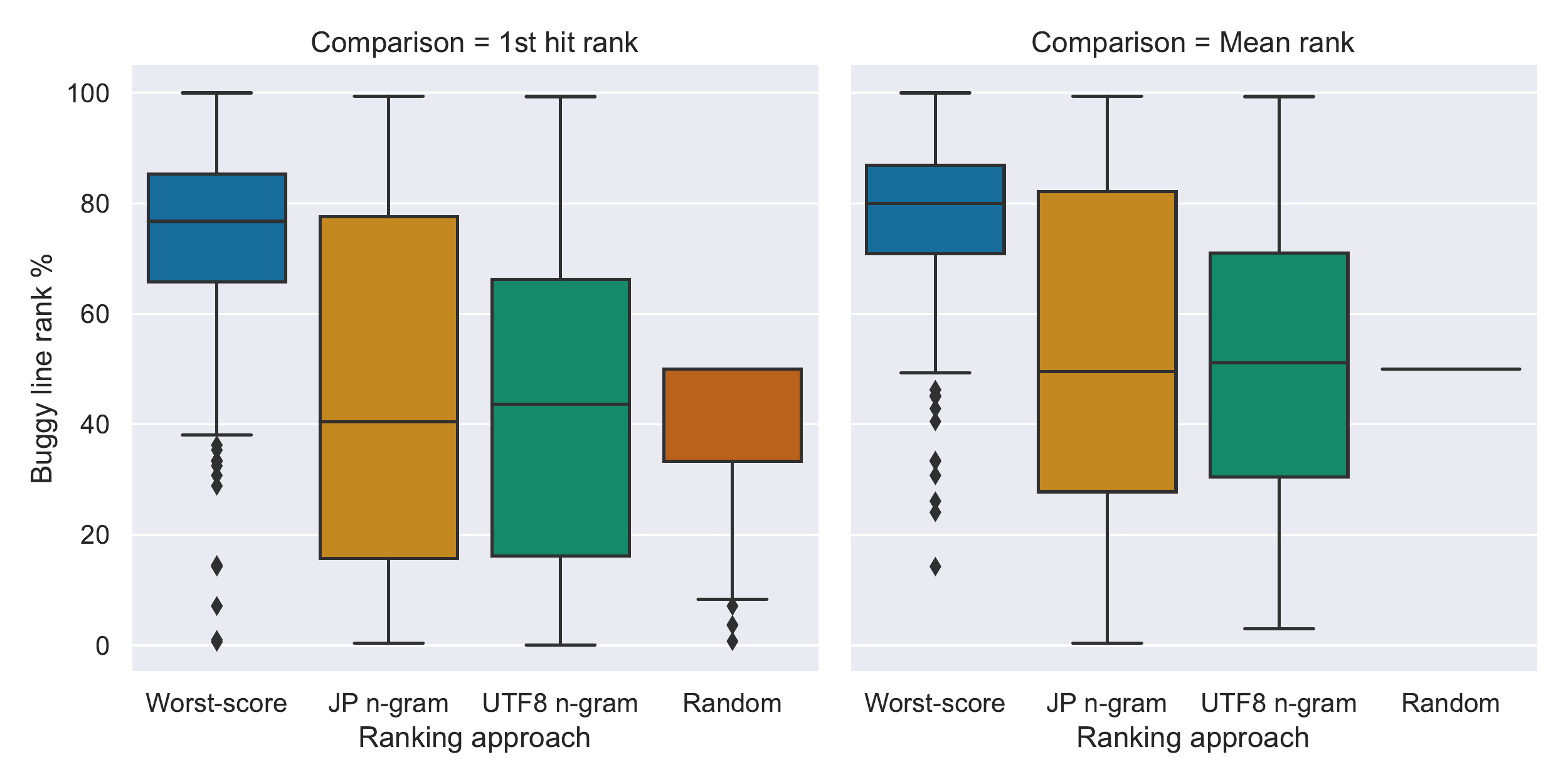}
    % \vspace{-3.5em}
    \caption{Comparison of the buggy lines rankings by worst possible scores, UTF8
n-gram and JP n-gram models (created respectively using UTF8 and
Java Parser tokenizers) when targeting the buggy versions not exposing any business-logic-related buggy line. n-gram techniques perform similarly to random on these subject buggy versions. %\toolname{} performs significantly worst than Random as it attributes the worst score to all not business-logic-related lines. 
}
    \label{fig:boxplot_ranking_not_covered_by_CB}
    %\vspace{-0.8em}
\end{figure}

To check whether this observed decrease of performance for n-gram is indeed caused by the additional lines or because they performed worst on the previously excluded bugs from our dataset, we reproduce the same comparison on the subset of our dataset where all bugs are located outside of the business-logic code, implying that %counting the bugs where \toolname{} did not cover any buggy-line (all of them are outside of the business-logic code), thus, attributing the worst score to all buggy lines. 
\toolname{} will attribute the worst score to every buggy line.
We illustrate the rankings distribution of the "worst-score" strategy (ranking all buggy-lines last), JP and UTF8 n-gram models and uniform-random in Figure~\ref{fig:boxplot_ranking_not_covered_by_CB}.

Although small, the n-gram models kept some advantage over random ranking as in the Figures \changing{\ref{fig:RQ3}} and \changing{\ref{fig:RQ4}}, in contrast to \toolname{}'s "worst-score" results.% which dropped drastically. 
The contrasting results between the Figures~\ref{fig:boxplot_ranking_not_covered_by_CB} and \changing{\ref{fig:RQ4}} highlight the negative impact of ranking the not business-logic lines by naturalness as they compensated \toolname{}'s disadvantage of attributing the worst ranks to buggy lines, in \changing{10\%} of the studied cases. 
Consequently, these results reinforce our conclusion that including the non-business-logic lines in the analysis adds noise to the search-space~\cite{natSoftwareRevisited2019}, and consequently hinders the ranking accuracy.

%This supports the observation that these source-code lines (i.e. the imports, the brackets, etc.) do not present any distinguishable not-natural symptom even when the code is buggy.

%the business-logic  by the fact that CodeBERT has a design-disadvantage in this experiment, considering that it did not rank any buggy line for around 10\% of the studied bugs, which lead to very low rankings. We illustrate 

%To verify this finding, we extend this study by considering another version of n-gram ranking that favours first the business-logic lines over the remaining ones before sorting both subsets by their cross-entropies.

%The comparative results of this sorting combinations is illustrated in Figure \changing{\ref{fig:RQ5}}.

%\ahmed{add a boxplot and check this out.}

\subsection{Which metric to use for which bug?}

%\begin{figure}[t]
%\centering 
 %    \begin{subfigure}{0.4\textwidth}
  %       \centering
   %      \includegraphics[width=\textwidth]{figurestables/venn_rq2__p_1h.pdf}
    %    \vspace{-0.5em}
     %    \caption{1st hit rank: 1st ranked buggy line.}
      %   \label{fig:venn_p1h}
%     \end{subfigure}
 %    \hfill
  %   \begin{subfigure}[b]{0.4\textwidth}
   %      \centering
    %     \includegraphics[width=\textwidth]{figurestables/venn_rq2__p_m.pdf}
     %    \caption{Mean rank: average rank of buggy lines.}
      %   \label{fig:venn_m}
%     \end{subfigure}
 %    \hfill
  %   \caption{which metric ranks the best the buggy lines. \textbf{analysis:} CodeBERT confidence performs the best for around 50\% of the cases, followed by cosine then matchorig which perms almost similarly to random. \textbf{conclusion:}  1) no big benefit in using matchorig 2) it could be interesting to complement confidence, using cosine.}
%    \label{fig:venn}
%\end{figure}

\begin{figure}[t]
\centering 
    %\vspace{-1.0em}
    %\adjincludegraphics[width=0.5\textwidth]{figurestables/venn_rq2__p_m.pdf}
    \adjincludegraphics[height=7cm,trim={{.1\width} {.2\width} {.1\width} {.1\width}},clip]{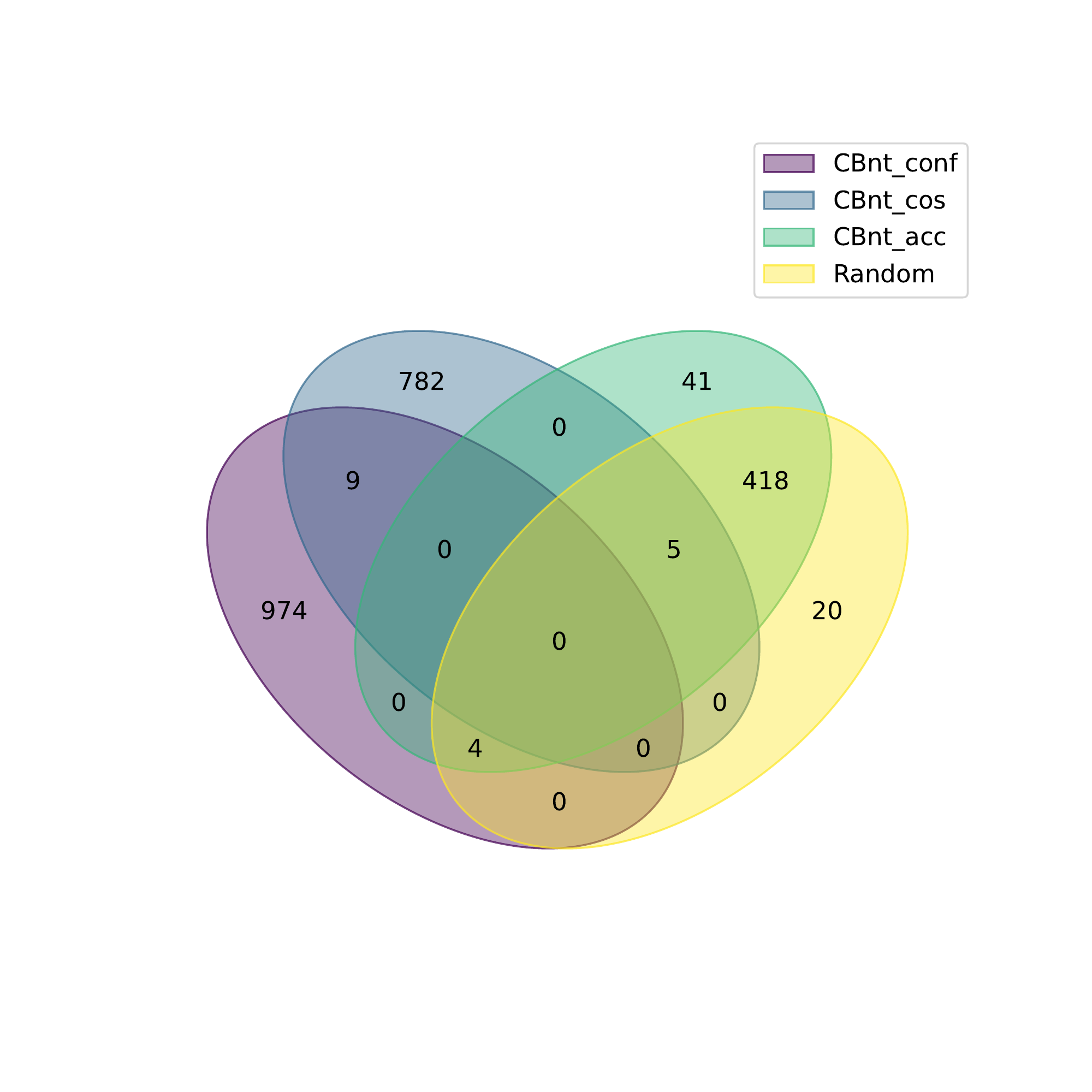}
    % \vspace{-3.5em}
    \caption{Which metric ranks the best the buggy lines, in most of the cases? CodeBERT confidence \conf performs the best for around 50\% of the cases, followed by \cosin then \acc which perms almost similarly to random. Therefore, There's no big benefit in using \acc while it could be interesting to complement \conf capabilities, using \cosin.}
    \label{fig:venn}
\end{figure}

\begin{figure*}[t]
\centering
\vspace{-0.6em}
\begin{subfigure}{0.26\textwidth}
%\includegraphics[width=\textwidth]
% left , bottom , right , up
\adjincludegraphics[width=\textwidth, trim={{.26\width} {.05\width} {.1\width} {.1\width}} ,clip]{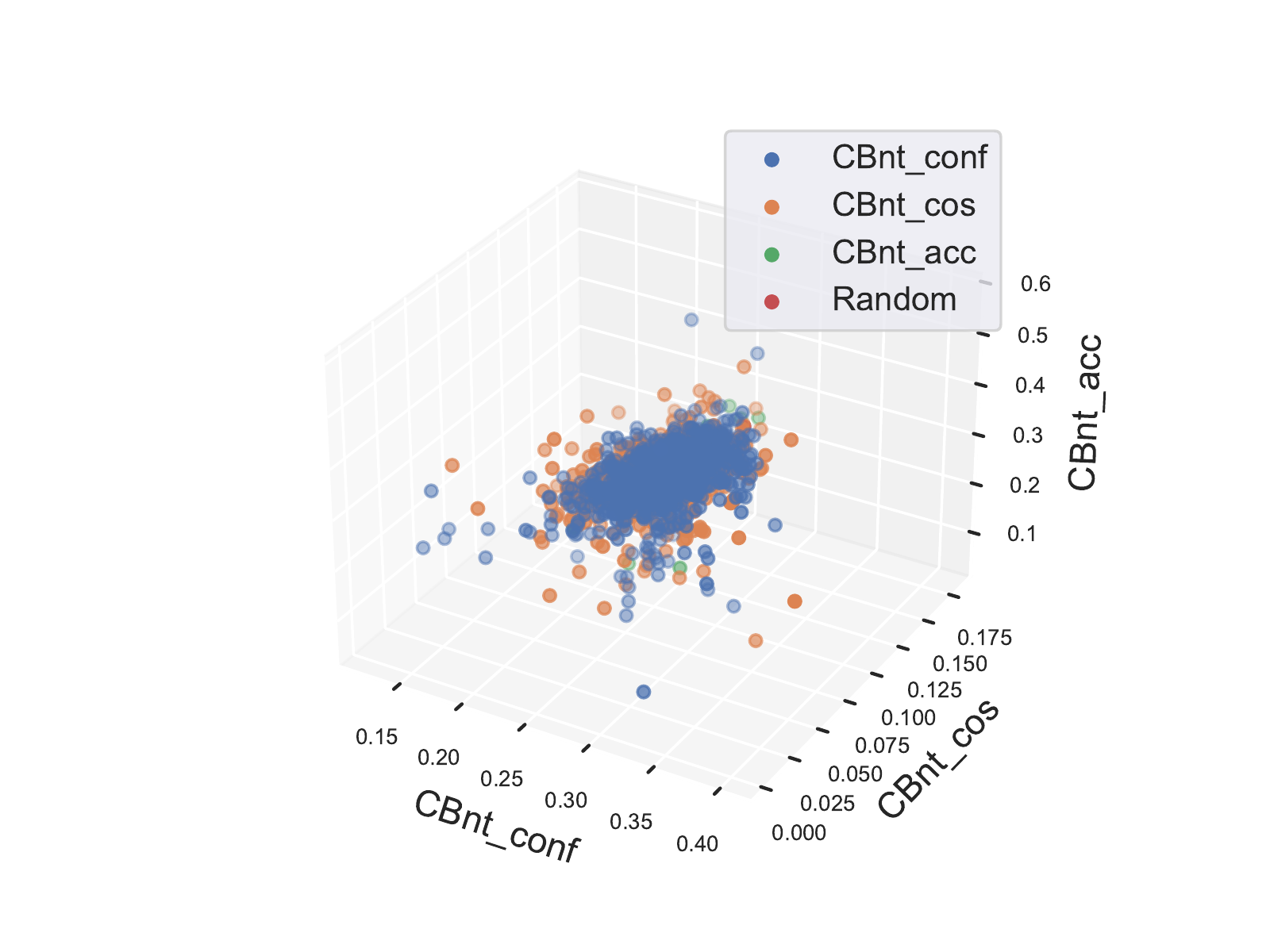}
%\vspace{-2.0em}
\caption{{\small Scatterplot of bugs best-ranking-metric by standard deviation.}}
\label{fig:sd_3d}
\end{subfigure}
\begin{subfigure}{0.73\textwidth}
\centering
\adjincludegraphics[width=\textwidth, trim={{0.01\width} {.01\width} {.04\width} {.01\width}} ,clip]{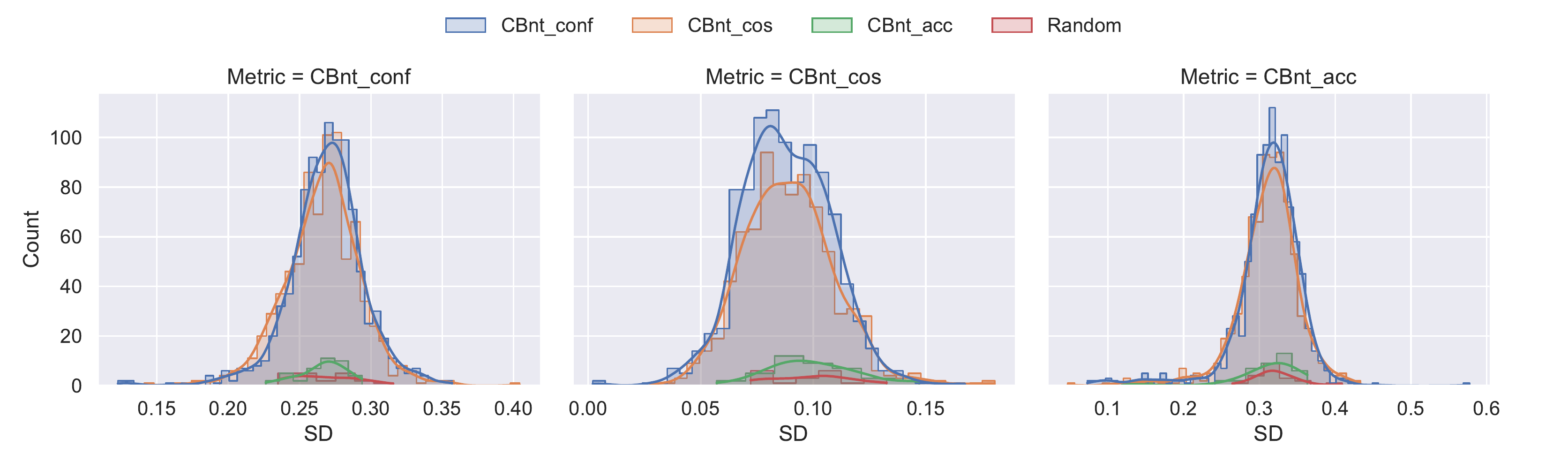}
%\vspace{-2.0em}
\caption{{\small Histogram of bugs best-ranking-metric by standard deviation.}}
\label{fig:sd_hist}
\end{subfigure}
\caption{Distribution of Bugs best-ranking-metric by standard deviation of the metrics measured on their corresponding subject lines. Except for few scores, \conf ranks the majority of the bugs the best independently from the measured SDs. %: every buggy-version is represented by a dot whose coordinates correspond to its standard-deviation of the measured metrics $(SD($\conf{}$),SD($\cosin{}$),SD($\acc{}$))$. 
%We cannot distinguish which metric ranks the buggy lines of a considered set of by lines from the SD of its scores.
}.
\vspace{-1.6em}
\label{fig:sd}
\end{figure*}

% \begin{figure}[t]
% \centering 
%      \begin{subfigure}{0.5\textwidth}
%          \centering
%          \includegraphics[width=\textwidth]{figurestables/boxplot_sd_p_1h.pdf}
%         \vspace{-0.5em}
%          \caption{1st hit rank: 1st ranked buggy line.}
%          \label{fig:sd_1h}
%      \end{subfigure}
%      \hfill
%      \begin{subfigure}[b]{0.5\textwidth}
%          \centering
%          \includegraphics[width=\textwidth]{figurestables/boxplot_sd_p_m.pdf}
%          \caption{Mean rank: average rank of buggy lines.}
%          \label{fig:sd_m}
%      \end{subfigure}
%      \hfill
%      \caption{standard deviation of the considered metrics does not help at all: most of the bugs where confidence (the metric to complement) performs worst than the other metrics have a SD within the range of SD corresponding to the bugs better ranked by confidence.}
%     \label{fig:sd}
% \end{figure}

The empirical results driven on our large set of buggy versions highlight the effectiveness of \toolname{} in capturing the naturalness of source-code, especially via its low-confidence metric.
However, from the outliers in the Figures~\ref{fig:RQ1a}, \ref{fig:RQ2}, \ref{fig:RQ3} and~\ref{fig:RQ4}, we notice that \toolname{} does not perform equally on all considered bugs.
Which means for instance that it outperforms uniform-random ranking in the majority of the cases, but yields worst rankings for a small portion of the dataset.
Therefore, we turn our attention towards investigating the possibility of better handling those bugs, leveraging one of the other \toolname{} metrics. 

We start by mapping every metric with the bugs on which it attributed the best mean ranking to the buggy lines. 
In figure~\ref{fig:venn}, we illustrate a Venn diagram of this distribution, including uniform-random-ranking as baseline.  
%We illustrate the distribution of bugs grouped by whether a metric ranks them the best in the Figure~\ref{fig}. 
%To have a better vision on the impact of the analysed buggy-versions, and whether the embeddings cosine and prediction accuracy could amend the effectiveness of the low-confidence, we 
%study 
%the buggy distribution and the variance of the prediction metrics of the considered bugs.
As shown in our results, the \toolname{} confidence is the best naturalness indicator for the majority of bugs, followed by the cosine similarity and the prediction accuracy. 
We also notice that except for a minority of \changing{20} bugs, at least one of the \toolname{} metrics achieves better scores or similar to random-uniform ranking. 
Additionally, except the large intersection set of bugs that are best-ranked by the prediction accuracy and uniform-random-ranking, the metrics rarely achieve their best rankings of the buggy lines for the same bugs.
This observation introduces the hypothesis that the metrics are complementary and eventually, one could rely on different metrics for different~bugs.

Aiming at distinguishing between our dataset bugs, we measure the metrics standard deviation (SD) per %every one's 
studied lines. 
In this setup, we exclude from our clusters the bugs that are intersecting with random and plot the SD distributions of the remaining ones, in Figure~\changing{\ref{fig:sd}}.
% extracting %  with the aim to define which metric performs the best for a which lines set.  
%Therefore, we seek at investigating the possibility of amending the capability of \toolname{}'s ranking by using a selective combination of its metrics. In other words, we rank the lines by \conf{}, whenever this metric is performing the best, and by one of the other metrics in the other cases.
%Intuitively, we study these metrics variation per buggy-version-lines among each cluster, with the aim to define which metric performs the best for a given lines set.      
%In Figure~\changing{\ref{fig:sd}} we illustrate the distribution of the standard deviation (SD) of the each cluster related line-score-sets, in terms of attributing the lowest ranks to the buggy lines (mean ranks comparison). 
The plots illustrate that the bugs from different clusters share similar ranges of SD with mean values around \changing{0.27}, \changing{0.09} and \changing{0.31} for respectively \conf{}, \cosin{} and \acc{}.
Also in the majority of the cases, the SD of the bugs best treated by other metrics than \conf{} fall in the same range of this latter, thus, cannot be distinguished from each other. 
However, we notice that for some SD values, \cosin{} ranks better more bugs than \conf{}. This difference is noticeable for roughly: $SD(\conf{})$ values between \changing{0.2} and \changing{0.24} or above \changing{0.35}, $SD(\cosin{})$ values above \changing{0.14} and $SD(\acc{})$ values below \changing{0.05} or between \changing{0.19} and \changing{0.28} or between \changing{0.395 and 0.42}.
%However, from the outliers, we notice that the SD can guide us towards using the best metric. For instance, in the studied dataset, the confidence is never the best ranking metric for the bugs yielding SDs in the following cases:
%\begin{itemize}
 %   \item lines with very low SD in terms of prediction accuracy (\changing{$SD<0,05$}) are better ranked by uniform-random,
  %  \item lines with very high SD in terms of prediction confidence (\changing{$SD>0,38$}) are better ranked by embeddings-similarity
   % \item and lines with very high SD in terms embeddings-similarity (\changing{$SD>0,19$}) are better ranked by embeddings-similarity.
%\end{itemize}
These SD ranges represent a small fraction of our dataset, exactly \changing{357} bugs, among which \changing{193} where \cosin{} performed the best and \changing{200} where it outperformed \conf{}, which correspond respectively to \changing{15.8\%, 8.5\% and 8.8\%} of the studied bugs.
Nevertheless, these results may motivate future investigation on the use of \cosin{} over \conf{} in similar cases, on different setups. 
%,where the \toolname{}'s performance is boosted by \changing{XX\%}. 
%Moreover, this advantage is further emphasised by the calculated $\hat{\text{A}}_{12}$ ratios between both metrics' results in favour of \cosin{}, of \changing{XX} and \changing{XX} in terms of attributing respectively the lowest 1st buggy line rank and the mean buggy-lines ranking. 
%This is further emphase
%bring an advantage in a very small portion of our dataset, however open the space for further investigation on how we can combine the metrics depending on the variance of the features extracted from the studied source-code.

%\ahmed{Probably create a venn diagram of where the tools perform better than random and not better than each other.}

%\ahmed{fix the venn diagram: some bugs are missing because complexity was included in it - get rid of complexity from the dataframe...then fix the SD boxplot and use only the mean.}

%\ahmed{I moved this from RQ1 to discussion, because 1) RQ1 started to be crowded and 2) we observe a different trend for the accuracy than cosine and confidence.}

\subsection{Impact of generating more predictions per token?}
\label{subsec:pred-per-token}

\begin{figure}[t]
\centering 
    \vspace{-0.3em}
    \includegraphics[width=0.5\textwidth]{ 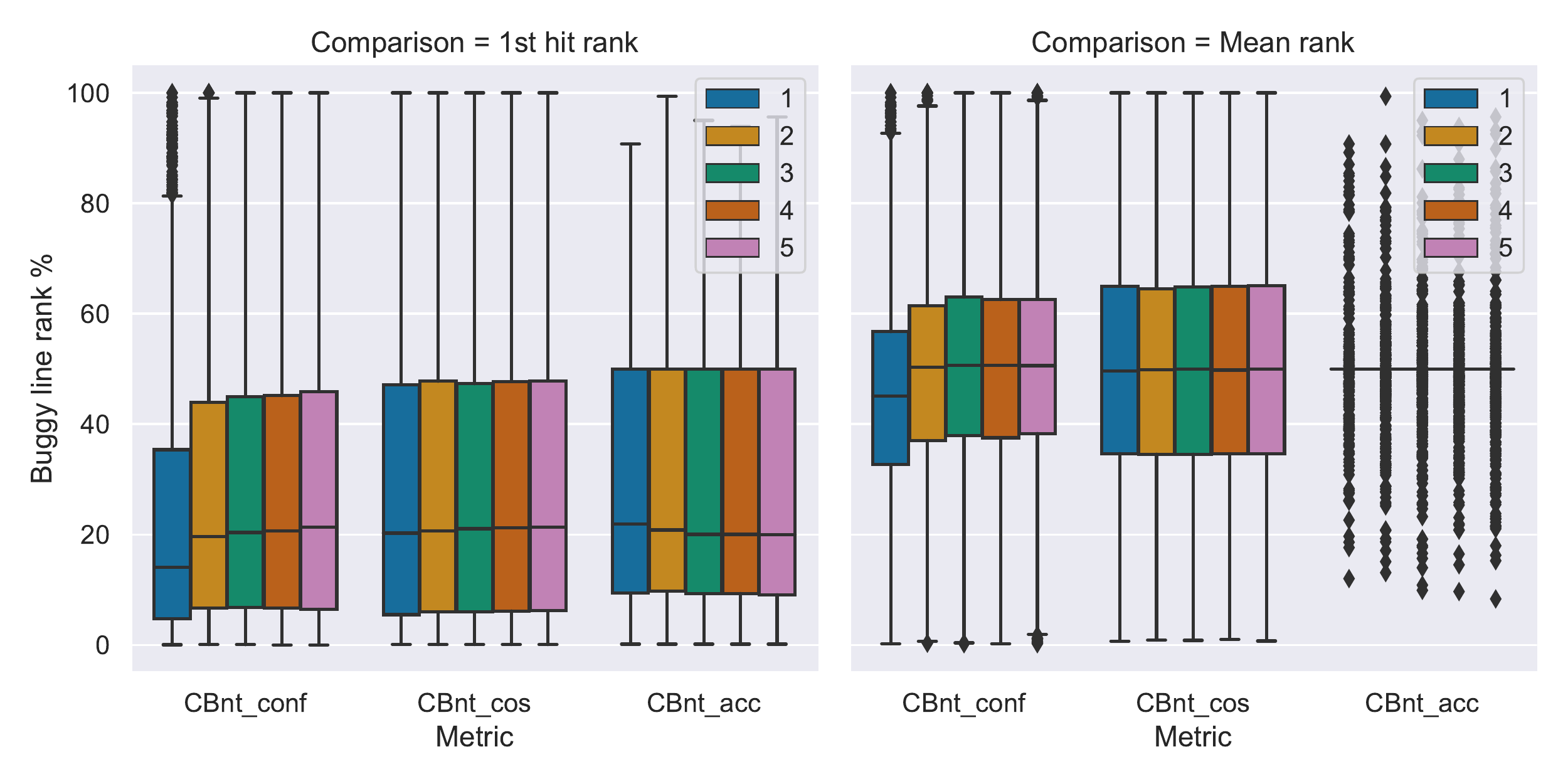}
    \vspace{-0.5em}
    \caption{Buggy lines ranking using 1, 2, 3, 4 and 5 predictions per token.
    The more predictions we use, the more the information about the confidence gets dissipated, thus the more the ranking performance decreases, except when considering the accuracy metric.}
    \label{fig:RQ1b}
    \vspace{-1.0em}
\end{figure}

% \begin{figure}[t]
% \centering 
%      \begin{subfigure}{0.5\textwidth}
%          \centering
%          \includegraphics[width=\textwidth]{figurestables/boxplot_df__rq1b1__p_1h_cbmupy.pdf}
%         \vspace{-0.5em}
%          \caption{1st hit rank: 1st ranked buggy line.}
%          \label{fig:RQ1b_1h}
%      \end{subfigure}
%      \hfill
%      \begin{subfigure}[b]{0.5\textwidth}
%          \centering
%          \includegraphics[width=\textwidth]{figurestables/boxplot_df__rq1b1__p_m_cbmupy.pdf}
%          \caption{Mean rank: mean rank of buggy lines.}
%          \label{fig:RQ1b_m}
%      \end{subfigure}
%      \hfill
%      \caption{Buggy lines ranking using 1, 2, 3, 4 and 5 predictions per token.
%     The more predictions we use, the more the information about the confidence gets dissipated, thus the more the ranking performance decreases, except when considering the accuracy metric.}
%     \label{fig:RQ1b}
% \end{figure}

To have a better understanding on whether generating more predictions from the model could improve the bugginess information retrieved by \toolname{}, we extend our experiment of RQ1 by comparing the ranking results using the best proven pairs of metric-aggregation from our results, when generating 1, 2, 3, 4 and 5 predictions per token. 
We illustrate in Figure~\ref{fig:RQ1b} the box-plots of the normalised rankings by number of lines of each bug.
Although comparable, the results depict a clear dissipation of the bugginess indicators retrieved from the prediction confidence and the cosine similarity metric, when we aggregate the values of more than one prediction by token.
In the other hand, we can see that the ranking performance %induced by the naturalness information extracted from 
effectuated by the prediction accuracy raises when we consider 2 and 3 predictions then converges to a stable value.
Besides the fact that this increase confirms further the correlation between code-naturalness and predictability, it remains negligible and keeps this metric-ranking far below the low-confidence one.
Therefore, we believe that it would be more cost-efficient and appealing for similar studies, to generate only 1 -- eventually up to 3 -- predictions instead of 5, as the default setting of CodeBERT.

%To have a better understanding of the impact of aggregating the scores of more predictions per token, on the buggy lines ranking, we have reproduced the same experiment by generating 2, 3, 4 and 5 predictions by token.
%To check whether generating more CodeBERT predictions by token has any impact on the ranking effectiveness, we compare the ranking results when aggregating the scores of 1, 2, 3, 4 and 5 predictions per token.  

%\ahmed{copy figure and conclusions fron RQ1 to here}

%% file: 9-conclusion.tex
\section{Conclusion}
\label{sec:conclusion}

Naturalness of code forms an important attribute often needed by researchers when building automated code analysis techniques. However, computing the naturalness of code using n-gram requires significant amount of work and a salable infrastructure that is not often available. An alternative solution is to use other readily available language models, perhaps more powerful than n-grams, such as transformer-based generative models (CodeBERT-like). Unfortunately, these models do not offer any token-based appearance estimations since they aim at generating tokens rather than computing their likelihood. To this end, we investigate the use of predictability metrics, of code tokens using the CodeBERT model and check their appropriateness in bug detection. Our results show that computing the confidence of the model when masking and generating a token, irrespective of whether the predicted token is the one that was actually predicted by the model, offers the best results, which are comparable (slightly better) to that of n-gram models trained on the code of the same project (intra-project predictions).